\documentstyle[aps,pre,epsfig]{revtex}
\newcommand{\figcite}[2]{(Taken from Figure #1 of Reference \cite{#2}).}
\newcommand{\mucw}{\hat{\eta}}
\newcommand{\pmat}[3]{\rho_{#1}^{#2}[#3]}
\newcommand{\pvec}[3]{p_{#1}^{#2}[#3]}

\newcommand{\dos}{G}
\newcommand{\hist}{H}
\newcommand{\equm}{0}
\newcommand{\Rset}[1]{\{\vec{R}\}_{#1}}
\newcommand{\rset}{\{\vec{r}\}}
\newcommand{\taumat}[1]{{\bf \tau}_{#1}}
\newcommand{\smat}{{\bf S}}
\newcommand{\uset}{\{u\}}
\newcommand{\qset}{\{q\}}
\newcommand{\qsetprime}{\{q^{\prime}\}}
\newcommand{\qsetref}[1]{\{q\}^{\mbox{\sc ref}}_{#1}}
\newcommand{\paraqset}{\{q\}^{(1)} \ldots \{q\}^{(\Omega)}}
\newcommand{\qsetalpha}{\{q\}_{\alpha}}
\newcommand{\lambdavalset}{\left[\lambda\right]}
\newcommand{\lambdaset}{\{\lambda\}}
\newcommand{\lambdasetprime}{\{\lambda^{\prime}\}}
\newcommand{\weight}{w}
\newcommand{\weightset}{\{w\}}
\newcommand{\deltapick}[2]{\Delta_{#1}[#2]}

\newcommand{\qsett} {\{q\}^{(t)} }
\newcommand{\duset}{\prod_i du_i }
\newcommand{\dqset}{\prod_i dq_i }

\newcommand{\plainop}{M}
\newcommand{\calop}{{\cal M}}
\newcommand{\plainopvalset}{\left [\plainop \right]}
\newcommand{\plainopvalsetalpha}{\left [\plainop \right]_\alpha}
\newcommand{\plainopset}{\left\{ \plainop \right\}}
\newcommand{\switchopphase}[3]{{#1}_{#2} (#3)}
\newcommand{\calfield}{\lambda}
\newcommand{\bigcon}{{\cal C}}
\newcommand{\bigconref}{\bigcon ^{\mbox{\sc ref}}}
\newcommand{\bigconset}{\{\bigcon \}}
\newcommand{\smallcon}{{\em c}}

\newcommand{\trialexpec}[1]{\langle #1 \rangle_{\trial}}
\newcommand{\expec}[2]{\langle #2 \rangle_{#1}}

\newcommand{\order}[1]{\mbox{O}[#1]}
\newcommand{\obs}{Q}
\newcommand{\eb}{\stackrel{eb}{=}}
\newcommand{\metrobare}[1]{{\mbox{min}\left\{ 1, #1\right\}}}
\newcommand{\metrofull}[1]{{\mbox{min}\left\{ 1, \exp\left[{-#1}\right]\right\}}}
\newcommand{\metrofunc}[1]{{\cal A}(#1)}
\newcommand{\metrofuncbare}{{\cal A}}
\newcommand{\pacc}{P_{\cal A}}
\newcommand{\samp}{{\cal S}}
\newcommand{\trial}{{\cal T}}
\newcommand{\ptrial}{P_{\cal T}}
\newcommand{\psamp}{P_{\samp}}
\newcommand{\psampprime}{P_{\samp^{\prime}}}
\newcommand{\pcanon}{P_0}
\newcommand{\gene}[2]{{\cal E} (#1 , \, #2)}
\newcommand{\genephase}[3]{{\cal E}_{#3} (#1 , \, #2)}
\newcommand{\genephasesimp}[2]{{\cal E}_{#2} (#1)}
\newcommand{\genesimp}{{\cal E}}

\newcommand{\genf}[2]{{\cal F}_{#1}(#2)}

\newcommand{\genpfsimp}{Z}
\newcommand{\genpfcalsimp}{{\cal Z}}
\newcommand{\genpf}[2]{Z_{#1}(#2)}
\newcommand{\alphatilde}{\tilde{\alpha}}
\newcommand{\intdqalpha}{\int \dqset \deltapick{\alpha}{\plainop}}
\newcommand{\intdq}{\int \dqset}
\newcommand{\intdu}{\int \duset}
\newcommand{\alphaalphatilde}{\alpha \alphatilde}
\newcommand{\PDF}{PDF}
\newcommand{\eqdot}{\stackrel{\cdot}{=}}

\begin{document}

\title{Computational strategies for mapping equilibrium phase diagrams}

\author{A.D. Bruce$^{(1)}$ and N.B. Wilding$^{(2)}$}

\address{$^{(1)}$Department of Physics and Astronomy, The University of Edinburgh\\
Edinburgh, EH9 3JZ,  United Kingdom\\[0.2cm]
$^{(2)}$Department of Physics, University of Bath\\
Bath, BA2 7AY, United Kingdom     
}

\maketitle

\begin{abstract}

We survey the portfolio of computational strategies available for
tackling the generic problems of phase behavior --
free-energy-estimation and coexistence-curve mapping.


\end{abstract}

\section{Introduction: defining our problem}

The phenomenon of {\em phase behavior} is generic to science; so we
should begin by defining it in a general way: it is the organization
of many-body systems into forms which reflect the interplay between
constraints imposed macroscopically (through the prevailing external
conditions) and microscopically (through the interactions between the
elementary constituents). In addition to its most familiar
manifestations in condensed matter science the phenomenon --and its
attendant challenges-- features in areas as diverse as gauge-theories
of sub-nuclear structure, the folding of proteins and the
self-organization of neural networks.

We will of course be rather more focused here.  We shall be concerned
with the {\em generic computational strategies} needed to address the
problems of phase behavior. The physical context we shall explore
will not extend beyond the {\em structural} organization of the elementary
phases (liquid, vapor, crystalline) of matter, although the {\em strategies} 
are much more widely applicable than this.
We shall have nothing
to say about a wide spectrum of techniques (density functional theory 
\cite{densfunc}, 
integral equation theories \cite{inteqn},
anharmonic perturbation theory \cite{anhpert},
virial expansions \cite{virial}) which have done
much to advance our understanding of sub-classes (solid-liquid,
solid-solid or liquid-gas) of phase behavior, but which are
less than `generic'. For the most part we shall also largely restrict 
ourselves to systems comprising simple classical particles in 
defect-free structures, interacting through a prescribed potential function.

\begin{figure}[h]
\epsfxsize=60mm
\begin{minipage}[t]{7cm}
\epsffile{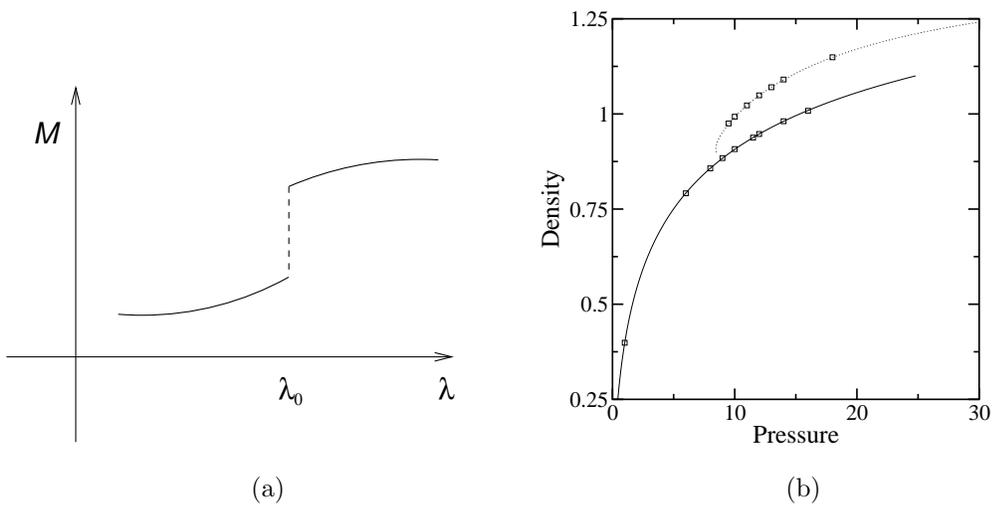}
\begin{center}
(a)
\end{center}
\end{minipage}
\begin{minipage}[t]{7cm}
\epsffile{figuredir/realexp.eps}
\begin{center}
(b)
\end{center}
\end{minipage}
\vspace*{1cm}

\caption{
(a) Schematic representation of the results of an `ideal experiment'
 on phase behavior, which may take as much time as we need: the equilibrium 
value of some physical quantity $\plainop$ changes discontinuously at some 
sharply-defined value, ${\calfield}_0$, of some field
$\calfield$.
(b) An example of the experimental reality: the results of a typical
simulation study of solid-liquid phase behavior of hard spheres; the measured 
density continues to follow the branch (liquid or solid) 
on which the simulation is initiated, well beyond the coexistence pressure
$P_x$ ($\simeq 11.3$ in these units) \protect  \cite{nbwalsounpublished}.
}

\label{fig:metastabhysteresis}

\end{figure}

There is one other rather more significant respect in which our
objectives are limited; we need to identify it now.  In its fullest
sense the `problem' of phase behavior entails questions of {\em kinetics}:
{\em how} phase transformations occur. Within a computational
framework such questions are naturally addressed with the techniques
of Molecular Dynamics (MD), which numerically integrate the equations
of motion associated with the many-body-potential, and thereby attempt
to replicate, authentically, the phase-transformation {\em process}
itself. There is a long, rich and distinguished history of activity of
this kind, providing insights into many (perhaps the most) challenging
and interesting issues associated with phase behavior. But the
`authenticity' carries a price: like their laboratory-counterparts,
such computer experiments display phenomena (hysteresis and metastability) associated with the long time-scales required for 
the reorganization processes. These phenomena are useful
qualitative signatures of a phase transformation and worthy of study
in their own right. But they tend to obscure the 
{\em intrinsically sharp} characteristics of the transformation 
(Fig.~\ref{fig:metastabhysteresis}).

If our interest is restricted to those sharp characteristics --in
particular {\em where} a phase transformation occurs in an `ideal
experiment'-- we may formulate (and in principle solve) the problem
entirely within the framework of the equilibrium statistical mechanics
of the competing phases (the framework which implies that `sharpness').

This is the stance we adopt here: we restrict our attention to the
task of mapping {\em equilibrium phase boundaries}. This choice leads
naturally to another.  Freed of the need to capture the authentic
dynamics, one finds that the natural computational
tool becomes Monte Carlo (MC) rather than MD. The strategic advantage
of this choice is the range of ways (not all of which have yet been
dreamt of) in which MC algorithms may be engineered to move around
configuration space. Indeed, understanding the distinctive features of
that `space' which are implied by the {\em occurrence} of phase
behavior, and engineering an algorithm (effectively a
pseudo-dynamics) to match, are the key themes of recent activities,
which we shall highlight in this work.

We shall begin (Section~\ref{SEC:BASICEQUIPMENT}) by assembling the
basic equipment. Section~\ref{subsec:formulations} formulates the
problem in the complementary languages of thermodynamics and
statistical mechanics.  The shift in perspective --from `free
energies' in the former to `probabilities' in the latter-- helps to
show what the {\em core problem} of phase behavior really is: a comparison of the {\it a priori} probabilities of two regions of configuration
space. Section~\ref{subsec:tools} outlines the standard portfolio of
MC tools and explains why they are not equal to the challenge posed by 
this core problem.

Section~\ref{SEC:PATHS} identifies and explores what proves to be {\em
the} key concept here - the notion of a {\em path through
configuration space}. Most approaches to the core problem utilize a
`path' which, in some sense, `connects' the two regions of
interest. They can be classified, helpfully we think, on the basis of
the choices made in regard to that path: the {\em route} it follows
through configuration space; and the way in which one contrives to
{\em sample} the configurations along that route. There are a
limited number of generically-distinct path-routing options; we
summarize them in Section~\ref{subsec:pathrouting}.
In Section ~\ref{subsec:pathsampling} we identify the
generically distinct path-sampling strategies. There are actually
fewer than one might suppose; this is an area in which many wheels
have been reinvented. Some of the new wheels at least run more quickly
than their precursors; the advances here
(the refinement of `extended sampling' techniques) 
account for much of the recent activity in this area. They
are essential for would-be practitioners but not for the logic of the story.
Accordingly they are addressed separately, in an Appendix.

The relatively few generic choices in regard to routing and sampling have been
deployed in combinations too numerous to mention let alone discuss. We
have, instead, elected to survey a small number of what we shall
call {\em path-based techniques} (combinations of routing and sampling
strategies) in some detail. This survey
(Section~\ref{SEC:PATHBASEDTECHNIQUES}) extends from the standard
technique of long-standing, numerical integration to reference macrostates,
to one only recently introduced, in which the core problem is solved
by a MC leap directly between the phases of interest.

We have chosen to organize our strategic survey around the
concept of a path. There are some strategies which do not naturally
fit into this framework, but which must certainly be included here: we
discuss them in Section~\ref{SEC:DOINGWITHOUTPATHS}.

Solving the core problem at some state point is the major part of the
task of determining the phase-boundary; but not quite all. To complete
the job one has to {\em track} a phase boundary (once one has found a
point on it) and one has to {\em extrapolate} from the limited sizes
of system that simulations can handle to the thermodynamic limit of
interest.  Section~\ref{SEC:RESTOFTHEJOB} reviews the complementary
tools needed.

We have chosen for the most part to focus on relatively idealised
model systems; in Section~\ref{SEC:IMPERFECTION} we consider some of
the issues associated with departures from the ideal.

Finally, section~\ref{SEC:OUTLOOK} offers some thoughts on how we are likely to 
make further progress.

There are three further caveats about what the reader can hope to find
in what follows. It is not easy (perhaps not possible) to be
exhaustive and clear; we aim to be clear. One can adopt an
organizational structure founded either on the history or the logic of
the ideas; we have chosen the latter. We have surely devoted
disproportionately much space to our own contributions; this is mainly
because we can explain them best.

\section{Basic equipment \label{SEC:BASICEQUIPMENT}}

\subsection{Formulation: statistical mechanics \& thermodynamics
\label{subsec:formulations}}

We will formulate the problem simply but generally.  Consider a system
of structureless, classical particles, characterized macroscopically
by a set of thermodynamic coordinates (such as the temperature $T$) and
microscopically by a set of model parameters which prescribe their
interactions. The two sets of parameters play a strategically similar
role; it is therefore convenient to denote them, collectively, by a
single label, $\smallcon$ (for `conditions or `constraints' or `control
parameters' in thermodynamic-and-model space). 

We are fundamentally concerned with the phase behavior rooted in the
{\em spatial} organization of the particles and reflected in the
statistical behavior of their position coordinates $\vec{r}_i,
i=1,N$. The components of these coordinates are the principal members
of a set of {\em generalized coordinates} $\qset$ locating the system in
its {\em configuration space}. In some instances (dealing with fluid
phases) it is advantageous to work with ensembles in which particle
number $N$ or system volume $V$ is free to fluctuate; the coordinate
set $\qset$ is then extended accordingly (to include $N$ or $V$) and
the control parameters $\smallcon$ extended to include the
corresponding fields (chemical potential $\mu$ or pressure $P$).

The statistical behavior of interest is encapsulated in
the equilibrium probability density function $\pcanon(\qset|\smallcon)$.
This \PDF\ is
determined by an appropriate ensemble-dependent, 
dimensionless \cite{factorsofT}
configurational energy 
$\gene{\qset}{\smallcon}$. The relationship takes the generic form
\begin{equation}
\label{eq:gendistdef}
\pcanon(\qset|\smallcon) = \frac{1}
{\genpf{}{\smallcon}} e^{-\gene{\qset}{\smallcon}}
\end{equation}
where the normalizing prefactor (the partition function) is defined by
\begin{equation}
\label{eq:genpfdef}
\genpf{}{\smallcon} \equiv \intdq e^{-\gene{\qset}{\smallcon} } 
\end{equation}
Some small print should appear here; we shall come back to it.  The
different phases which the system displays will in general be
distinguished by the values of some  macroscopic property
loosely described as an {\em order parameter}. Thus, for example,
the density serves to distinguish a liquid from a vapor; a structure
factor distinguishes a liquid from a crystalline solid. 
A suitable order parameter, $\plainop$, 
allows us to associate with each phase, $\alpha$,
a corresponding portion $\qsetalpha$ of $\qset$-space. We write that statement 
concisely in the form:
\begin{equation}
\label{eq:qsetalphadef}
\qset \in \qsetalpha 
\hspace{0.25cm}\mbox {iff}\hspace{0.25cm}
\plainop (\qset) 
\in \plainopvalsetalpha
\end{equation}
where $\plainopvalsetalpha$ is the set of order parameter values consistent 
with phase $\alpha$. 
The partitioning of $\qset$-space into distinct regions
is the key feature of the core problem.

The equilibrium properties of a particular 
phase $\alpha$ follow from the conditional counterpart 
of Eq.~\ref{eq:gendistdef} 
\begin{equation}
\label{eq:partialdistdef}
\pcanon(\qset| \alpha,\smallcon) = \left\{ \begin{array}{ll}
\frac{1}{\genpf{\alpha}{\smallcon}} e^{-\gene{\qset}{\smallcon}} & \qset \in \qsetalpha\\
0& \mbox{otherwise}\\
\end{array}
\right .
\end{equation}
with
\begin{equation}
\label{eq:partialpfdef}
\genpf{\alpha}{\smallcon } \equiv \intdq 
\deltapick{\alpha}{\plainop (\qset)}  
e^{-\gene{\qset}{\smallcon}} \equiv e^{-\genf{\alpha}{\smallcon}}
\end{equation}
The last equation defines $\genf{\alpha}{\smallcon}$, the {\em free energy} 
of phase $\alpha$, while
\begin{equation}
\label{eq:pickphaseqset}
\deltapick{\alpha}{\plainop} \equiv 
\left\{ 
\begin{array}{ll}
1 &  \plainop \in \plainopvalsetalpha
\\
0 & \mbox{otherwise}
\end{array}
\right.
\end{equation}
so that the integral is effectively confined
to the set of configurations $\qsetalpha$ associated with phase $\alpha$.

Since the notation does not always make it apparent, we should note that 
$\alpha$ and $\smallcon$
play operationally similar roles as {\em macrostate labels}: together they identify the
distinct sets of equilibrium macroscopic properties emerging from the equilibrium distribution, Eq.~\ref{eq:gendistdef}.
For some purposes we will find it useful to concatenate the two labels into a single `grand' macrostate label 
\begin{equation}
\label{eq:bigcon}
\alpha , \smallcon \longrightarrow \bigcon
\end{equation}
For the time being we shall continue to display the two labels separately.

Now appealing  back to Eq.~\ref{eq:gendistdef} we may write 
\[
\genpf{\alpha}{\smallcon} =\intdqalpha e^{-\gene{\qset}{\smallcon}} 
=\genpf{}{\smallcon} \intdqalpha \pcanon(\qset|\smallcon) 
\]
The {\em a priori} probability of phase $\alpha$ may thus be related 
to its free energy by
\begin{equation}
\label{eq:palphatofe}
\pcanon(\alpha|\smallcon) \equiv \intdqalpha \pcanon(\qset|\smallcon) 
=\frac{\genpf{\alpha}{\smallcon}}{\genpf{}{\smallcon}}
=\frac{e^{-\genf{\alpha}{\smallcon}}}{\genpf{}{\smallcon}}
\end{equation}
Alternatively
\begin{equation}
\label{eq:palphatopdf}
\pcanon(\alpha|\smallcon) \equiv \intdqalpha \pcanon(\qset|\smallcon) 
=\int d \plainop \deltapick{\alpha}{M} \pcanon(\plainop |\smallcon) 
\end{equation}
where $\pcanon(\cal M |\smallcon)$ is the equilibrium distribution of the 
chosen order parameter. Though seemingly naught but a tautology this 
representation proves remarkably fruitful

For two phases, $\alpha$ and $\alphatilde$ say, it then follows that
\begin{equation}
\label{eq:deltaf}
\Delta \genf{\alphaalphatilde}{\smallcon}
\equiv\genf{\alpha}{\smallcon}- \genf{\alphatilde}{\smallcon}
=\ln \frac{\genpf{\alphatilde}{\smallcon}}{\genpf{\alpha} {\smallcon}}
=\ln \frac{\pcanon(\alphatilde|\smallcon)}{\pcanon(\alpha|\smallcon)}
=\ln \frac{
\int d{\plainop} \deltapick{\alphatilde}{M} \pcanon(\plainop |\smallcon) 
}{
\int d{\plainop} \deltapick{\alpha}{M} \pcanon(\plainop |\smallcon) 
}
\end{equation}

This is a key equation in several respects: it is conceptually helpful;
it is cautionary; and it is suggestive, strategically.

At a conceptual level, Eq.~\ref{eq:deltaf} provides a helpful link
between the languages of thermodynamics and statistical mechanics.
According to the familiar mantra of thermodynamics the favored phase
will be that of {\em minimal free energy}; from a statistical
mechanics perspective the favored phase is the one of {\em maximal
probability},
\underline{given} the probability partitioning implied 
by Eq.~\ref{eq:gendistdef}.

We also then see (the `cautionary' bit) that the thermodynamic mantra
presupposes the validity of Eq.~\ref{eq:gendistdef}, and thence of its
`small print', which we must now spell out.  In general
Eq.~\ref{eq:gendistdef} presupposes ergodicity on the space
$\qset$. The framework can thus be trusted to tell us what we will
`see' for some given $\smallcon$ (the `favored phase') only to the
extent that appropriate kinetic pathways exist to allow sampling
(ultimately, {\em comparison}) of the distinct regions of
configuration space associated with the different phases. In the
context of laboratory experiments on real systems the relevant
pathways typically entail the nucleation and growth of droplets of one
phase embedded in the other; the associated time scales are long; and
Eq~\ref{eq:deltaf} will be relevant only if the measurements extend
over correspondingly long times. The fact that they frequently do not
is signaled in the phenomena of metastability and hysteresis we have
already touched on.

Finally, Eq.~\ref{eq:deltaf} helps to shape strategic thinking on how
to broach the problem computationally.

It reminds us that what is relevant here is the {\em difference}
between free energies of two competing phases and that this free
energy difference is a {\em ratio} of the {\em a priori} probabilities
of the two phases. It implies that the phase boundary may be
identified as the locus of points of equal {\em a priori} probability
of the two phases, and that such points are in principle identifiable
through the condition that the order parameter distribution will have
equal integrated weights (areas) in the two phases. The discussion of
the preceding section also suggests that the pathways by which our
simulated system passes between the two regions of configuration space
associated with the two phases will play a strategically crucial
role. While, for the reasons just discussed, the details of those
pathways are essential to the {\em physical applicability} 
of Eq.~\ref{eq:deltaf}
they are {\em irrelevant} to the {\em values} of the quantities it
defines; we are thus free to engineer whatever pathways we may wish.

These considerations lead one naturally to the Monte Carlo toolkit.

\subsection{Tools: elements of Monte Carlo \label{subsec:tools}}

The Monte Carlo method probably ranks as the most versatile
theoretical tool available for the exploration of many-body
systems. It has been the subject both of general pedagogical texts
\cite{mcpedagogy} and applications-focused reviews
\cite{mcreviews}. Here we provide only its elements - enough to
understand why, if implemented in its most familiar form, it does not
deliver what we need, and to hint at the extended framework needed to
make it do so.

The MC method generates a sequence (Markov chain) of configurations in
$\qset$-space. The procedure can be constructed to ensure that, in
the `long-enough-term', configurations will appear in that chain with
{\em any} probability density, $\psamp(\qset)$ (the `S' stands for 
`sampling') we care to nominate. 
The key requirement (it is not strictly necessary \cite{dbnotnecessary}; 
and --as we shall see-- it is not always sufficient) 
is that the transitions, from one configuration  $\qset$ to another $\qsetprime$, should respect
the detailed balance condition
\begin{equation}
\label{eq:detailedbalance}
\psamp (\qset)
\psamp (\qset \rightarrow \qsetprime)
=
\psamp (\qsetprime)
\psamp (\qsetprime \rightarrow \qset)
\end{equation}
where 
$\psamp(\qset \rightarrow \qsetprime)$
is the transition probability, the probability density of
configuration $\qsetprime$ at Markov chain step $t+1$ given
configuration $\qset$ at time $t$. (We have added a subscript 
to emphasize that its form is circumscribed by the choice of 
sampling density, through Eq.~\ref{eq:detailedbalance}.)
MC transitions satisfying this constraint are realized
in a two-stage process. In the first stage, 
one generates a trial configuration $\qsetprime =\trial \qset$, where 
$\trial$ is some generally stochastic selection procedure; the 
probability density of a trial configuration
$\qsetprime$ given $\qset$ is of the form
\begin{equation}
\label{eq:ptrial}
\ptrial (\qsetprime \mid \qset)
=
\trialexpec{\delta(\qsetprime - \trial \qset)}
\end{equation}
where $\trialexpec{\cdot}$ represents an average with respect to the 
stochastic variables implicit in the procedure $\trial$.
In the second stage
the `trial' configuration is accepted (the system `moves' from  
$\qset$ to $\qsetprime$ in configuration space) with
probability $\pacc$, 
and is otherwise rejected (so the system `stays'  at $\qset$);
the form of the 
acceptance probability is prescribed by our choices for 
$\psamp$ and $\ptrial$ since 
\[
\psamp (\qset \rightarrow \qsetprime)
=
\ptrial (\qsetprime \mid \qset)
\pacc   (\qset \rightarrow \qsetprime)
\]
It is then easy to verify that the detailed balance condition
(Eq.~\ref{eq:detailedbalance}) is satisfied, if the acceptance 
probability is chosen as
\begin{equation}
\label{eq:metropacc}
\pacc (\qset \rightarrow \qsetprime) = 
\metrobare{
\frac
{\psamp(\qsetprime) \ptrial (\qset \mid \qsetprime)}
{\psamp(\qset) \ptrial (\qsetprime \mid \qset)}
}
\end{equation}
Suppose that, in this way, we build a Markov chain comprising a total of $t_T$ steps; we set aside the first 
$t_E$ configurations visited; we denote by
$\qsett$ ($t=1,\ldots t_U$) the configurations associated with the subsequent
$t_U\equiv t_T-t_E$ steps. The promise on the MC package is that
the expectation value $\expec{\samp}{\obs}$ of some observable 
$\obs$ = $\obs (\qset)$ {\em defined} by
\begin{equation}
\label{eq:sampexpecvalue}
\expec{\samp}{\obs} = \intdq \psamp (\qset) \obs (\qset)
\end{equation}
may be {\em estimated} by the sample average
\begin{equation}
\label{eq:estimator}
\expec{\samp}{\obs} \eb \frac{1}{t_U} \sum_{t=1}^{t_U} \obs (\qsett)
\end{equation}

Now we must consider the choices of $\psamp$ and $\ptrial$. 
Tailoring those choices to 
whatever task one has in hand provides potentially limitless
opportunity for ingenuity. 
But at this point we consider only the simplest possibilities.
The sampling distribution $\psamp(\qset)$ is chosen to be the appropriate 
{\em equilibrium} distribution $\pcanon(\qset \mid \smallcon)$
(Eq.~\ref{eq:gendistdef}) so that the configurations  
visited are representative of a `real' system, even though their 
sequence is not an authentic representation of the `real' dynamics.
We shall refer to this form of sampling 
distribution as {\em canonical}.
The trial-coordinate selection procedure $\trial$ is chosen to comprise
some {\em small} change of  {\em one} coordinate; the 
change is chosen to be small enough
to guarantee a reasonable acceptance probability (Eq.~\ref{eq:metropacc}), 
but no smaller, or the Markov chain will wander unnecessarily slowly 
through the configuration space. 
We shall refer to this form of selection procedure  as {\em local}.
For such schemes (and sometimes for others) the selection probability
density typically has the symmetry:
\begin{equation}
\label{eq:ptrialsymmetry}
\ptrial (\qset\rightarrow \qsetprime) = \ptrial(\qsetprime \rightarrow \qset)
\end{equation}
With these choices,
Eq.~\ref{eq:metropacc} becomes
\begin{equation}
\label{eq:metropaccsimple}
\pacc (\qset \rightarrow \qsetprime) = 
\metrofunc{\Delta \genesimp}
\end{equation}
where
\begin{equation}
\label{eq:deltagenedef}
\Delta \genesimp  \equiv \gene{\qsetprime}{\smallcon}  - \gene{\qset}{\smallcon} 
\end{equation}
and
\begin{equation}
\label{eq:metrofuncdef}
\metrofunc{x} \equiv \metrofull{x}
\end{equation}
defines the Metropolis acceptance function\cite{metropolis}.

These choices are not only the simplest, they are also the most frequent:
the local-canonical  strategy is the staple 
Monte Carlo method, and
has contributed enormously to our knowledge of many-body systems.

From what we have said  it would seem that this staple strategy would also
deliver what we require here. If, as promised, the Markov chain visits
configurations with the canonical 
probability (Eq.~\ref{eq:gendistdef}) we should
merely have to determine 
\begin{equation}
\pcanon(\alpha|\smallcon) \equiv \expec{\equm}{\Delta _{\alpha}}
\label{eq:palphadirect}
\end{equation}
from its estimator (Eq.~\ref{eq:estimator}),
effectively the proportion of time
the system is found in  region $\qsetalpha$.
Eq.~\ref{eq:deltaf} would then take care of the rest.
But the local-canonical strategy fails us here. 
The Markov chain typically does not extend beyond
the particular region of configuration space $\qsetalpha$ in which it
is initiated and the distribution of the (any) `order parameter' will capture only
the contributions associated with that phase.
The observations thus provide no basis for assigning a value to
the relative probabilities of the two phases, and thus of estimating the location of the phase boundary. 

This failure is a reflection 
of {\em both} of our `simple' choices. Firstly, the {\em local} character of 
coordinate updating yields a dynamics which (though scarcely authentic)
shares the essential problematic feature of the kinetic pathways supported by `real' dynamics:
evolution from one phase to another will typically require a traverse
through intermediate regions in which the configurations have a two-phase 
character. Secondly, the choice of a {\em canonical} sampling
distribution ensures that the probability of such intermediate 
configurations is extremely 
small; inter-phase traverses occur 
only on correspondingly long time scales. At this point one remembers
the small-print (`in the long term') that accompanies the MC toolkit. 
As a result
the effective sampling distribution is not the nominal choice,
$P(\qset|\smallcon)$ (Eq.~\ref{eq:gendistdef}) but $P(\qset| \alpha,\smallcon)$
(Eq.~\ref{eq:partialdistdef}), with the phase $\alpha$ determined by our choice
of initial configuration.
Since the sampling distribution appears
only in the acceptance probability (Eq.~\ref{eq:metropacc}), and then
only as a {\em ratio}, the algorithm does not distinguish between
$P(\qset|\smallcon)$ and $P(\qset| \alpha,\smallcon)$ --as long as it is trapped
in $\qsetalpha$.

We conclude that the local-canonical MC strategy cannot directly
deliver the simultaneous comparison between two phases which
(Eq.~\ref{eq:deltaf} suggests) provides the most efficient resolution
of the phase-boundary problem: in most circumstances this strategy will 
simply explore a {\em single} phase. 
We must now ask whether we can get by
with two {\em separate} (but still local-canonical) `single-phase' simulations, each determining the free-energy (or, equivalently, partition function) of one phase (Eq.~\ref{eq:partialpfdef}).

Let us first be clear about the circumstances in which a `single
phase' simulation makes sense. The brief and loose answer is: when the
time (Markov chain length) $t_{e \alpha}$ typical of escape from phase
$\alpha$ is long compared to the time $t_{s \alpha}$ required for effective sampling
of the configuration space of that phase.  More fully, and a little
more formally: when there exists some $t_{s \alpha} < t_{e \alpha}$ 
such that the
configuration set $\qsetalpha$ defined by Eq.~\ref{eq:qsetalphadef} is
effectively equivalent to that defined by the condition
\begin{equation}
\label{eq:qsetalphadefalt}
\qset \in \qsetalpha 
\hspace{0.25cm}\mbox {iff}
\hspace{0.25cm}
\qset 
\mbox{ is reachable from }
\qset ^{0}_{\alpha} 
\mbox{ within time }
t_{s \alpha}
\end{equation}
In this formulation, the configurations in $\qsetalpha$ are identified as 
those that may be reached in a simulation of length $t_{s \alpha}$,
initiated from some configuration $\qset ^{0}_{\alpha}$ 
that is associated with phase $\alpha$ but is otherwise arbitrary. 
The equivalence of
Eqs.~\ref{eq:qsetalphadef} and  \ref{eq:qsetalphadefalt} is assumed (usually
tacitly) in all `single phase' simulations.

Assuming these conditions are fulfilled, our single-phase
local-canonical algorithm will allow us to estimate the single-phase
canonical expectation value of any `observable' $\obs$ defined on the
configurations $\qset$
\begin{equation}
\label{eq:canonexpecvalue}
\expec{\equm,\alpha}{\obs} = \intdq \pcanon (\qset|\alpha, \smallcon ) \obs (\qset)
\end{equation}
from the average (Eq.~\ref{eq:estimator}) over a sample of the
canonically-distributed configurations.  But the single-phase
partition function $\genpfsimp_{\alpha}$ 
{\em is not} \cite{wellnotnaturally} an `average over
canonically distributed configurations'; rather, it measures the {\em
total effective weight} of the configurations that contribute
significantly to such averages. One can no more deduce it from a
sample of single phase properties than one can infer the size of an
electorate from a sample of their opinions. Thus local-canonical MC fails to 
deliver the absolute single-phase free energy also.

To determine the relative stability of two phases under conditions
$\smallcon$ thus requires a MC framework that, in some sense, does {\em more}
than sample the equilibrium configurations appropriate to the (two)
$\smallcon$-macrostates. We have seen where there is room for maneuver --
in the choices we make in regard to $\psamp$ and $\ptrial$. 
The possibilities inherent in the latter are intuitively obvious: better
to find ways of bounding or leaping through configuration space than be
limited to the shuffle of local-updating. The fact that we have flexibility in
regard to the choice of sampling distribution is perhaps less obvious so it 
is worth recording the simple result which shows us that we do. 

Let $\psamp$ and $\psampprime$ be two arbitrary distributions of the 
coordinates $\qset$. Then the expectation values of some arbitrary 
observable $\obs$ with respect to the two distributions are formally
related by the identity
\begin{equation}
\label{eq:identitygen}
\expec{S^{\prime}}{\obs} = \intdq \psampprime (\qset) \obs (\qset)
= \intdq \psamp (\qset) \obs (\qset)\frac{\psampprime (\qset)}{\psamp(\qset)} 
=\expec{S}{\frac{\psampprime}{\psamp}\obs}
\end{equation}
Thus, in particular, we can --in principle-- determine
{\em canonical} expectation values from an ensemble defined by an 
{\em arbitrary} sampling distribution through the relationship
\begin{equation}
\label{eq:identity}
\expec{\equm}{\obs} =\expec{S}{\frac{\pcanon}{\psamp}\obs}
\end{equation}
We do not {\em have } to make the choice $\psamp=\pcanon$. 
The issue of what sampling distribution will be
{\em optimal} was addressed in the
earliest days of computer simulation \cite{fosdick}. 
The answer depends on the 
observable $\obs$. In general the `obvious' choice 
$\psamp=\pcanon$,  though not strictly optimal, is adequate. 
But for some observables the choice of a canonical sampling distribution is
{\em so} `sub-optimal' as to be useless 
\cite{badsampling}. The core problem we face here
has a habit of presenting us with such quantities:
we have already seen one example (Eq.~\ref{eq:palphadirect}) and  we shall see others.

\section{Paths \label{SEC:PATHS}}

There are many ways of motivating, constructing
and describing the kind of MC sampling strategy we need;
the core idea we shall appeal to here
to structure our discussion is that of a {\em path}.

\subsection{Meaning and specification \label{subsec:meaningandspec}}

For our purposes a {\em path} comprises a sequence of contiguous
macrostates, $\bigcon_1, \bigcon_2 ...\bigcon_{\Omega} \equiv \bigconset$ \cite{useofbigcon}.
By
`contiguous' we mean that each adjacent pair in the sequence ($\bigcon_j,
\bigcon_{j+1}$ say) have some configurations in common (or that a
configuration of one lies arbitrarily close to a configuration of the
other).  A path thus comprises a quasi-continuous band through
configuration space.

The physical quantities that distinguish the macrostates from one
another will fall into one or other of two categories, which we shall
loosely refer to as {\em fields}, and {\em macrovariables}. 
In the former category we include thermodynamic
fields, model parameters and, indeed, the conjugate \cite{meaningofconjugate} 
of any 
`macrovariable'. By `macrovariable' \cite{macrovariable}
we mean any collective property, aggregating 
contributions from all or large numbers of the constituent particles,
free to fluctuate in the chosen ensemble, but in general
sharply-prescribed, in accordance with the Central Limit Theorem. Note
that we do not restrict ourselves to quantities that feature on the
map of thermodynamics, nor to the parameter space of the physical
system itself: with simulation tools at our disposal there are
limitless varieties of parameters to vary and properties to observe. 

\subsection{Generic routes\label{subsec:pathrouting}}

\begin{figure}[h]
\epsfxsize=60mm

\begin{minipage}[t]{7cm}
\epsffile{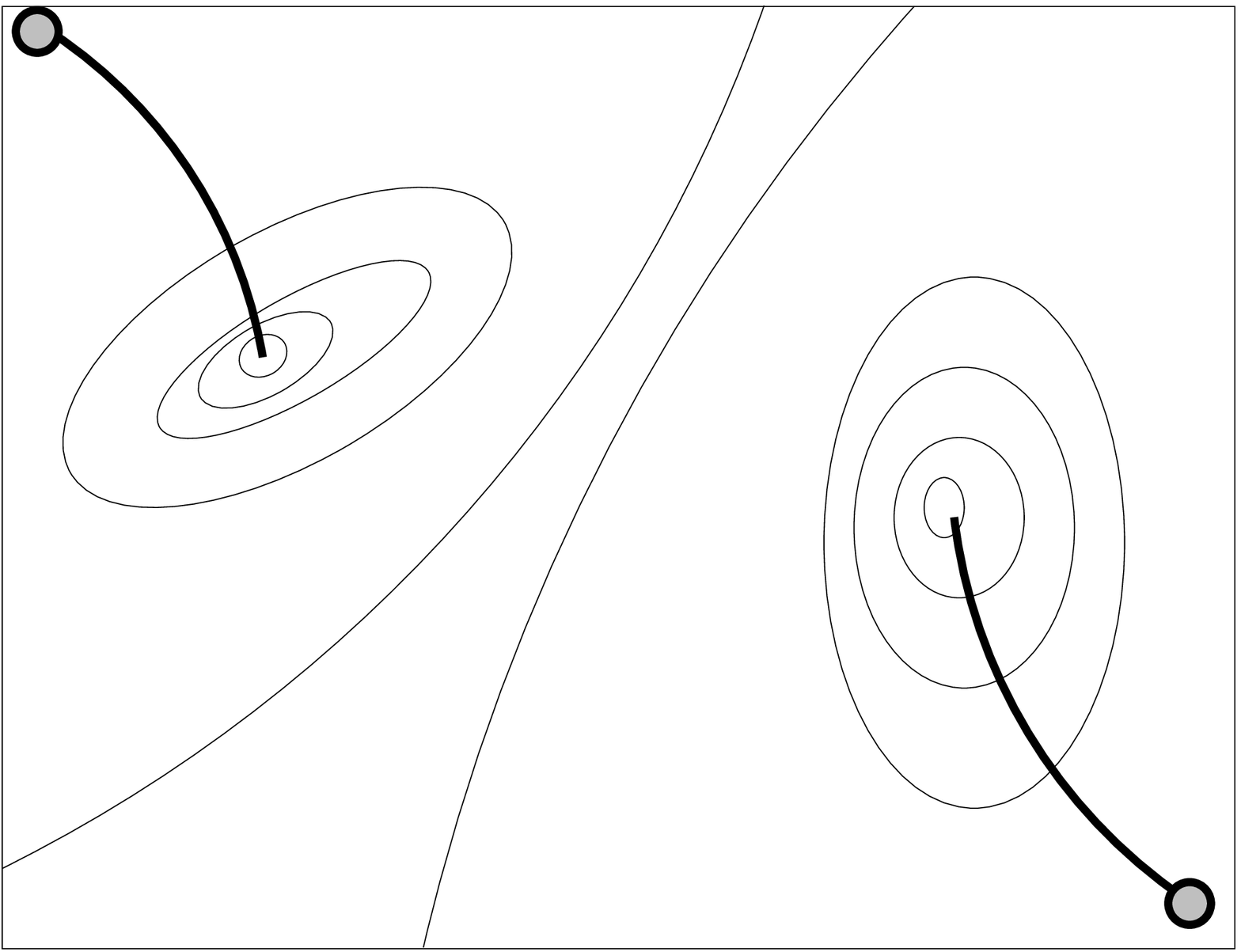}
\begin{center}
(a)
\end{center}
\end{minipage}
\begin{minipage}[t]{7cm}
\epsffile{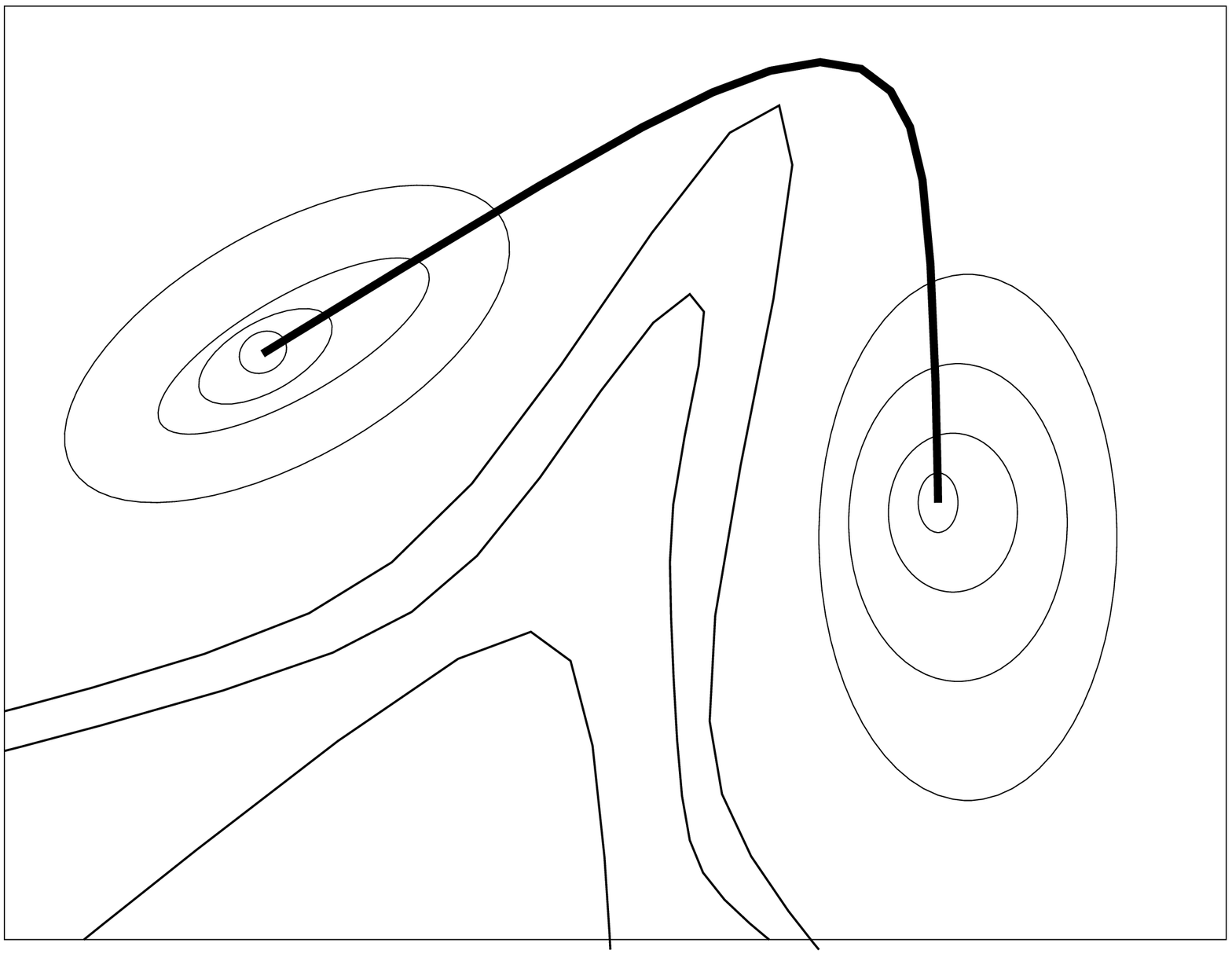}
\begin{center}
(b)
\end{center}
\end{minipage}

\vspace*{0.5cm}

\begin{minipage}[t]{7cm}
\epsffile{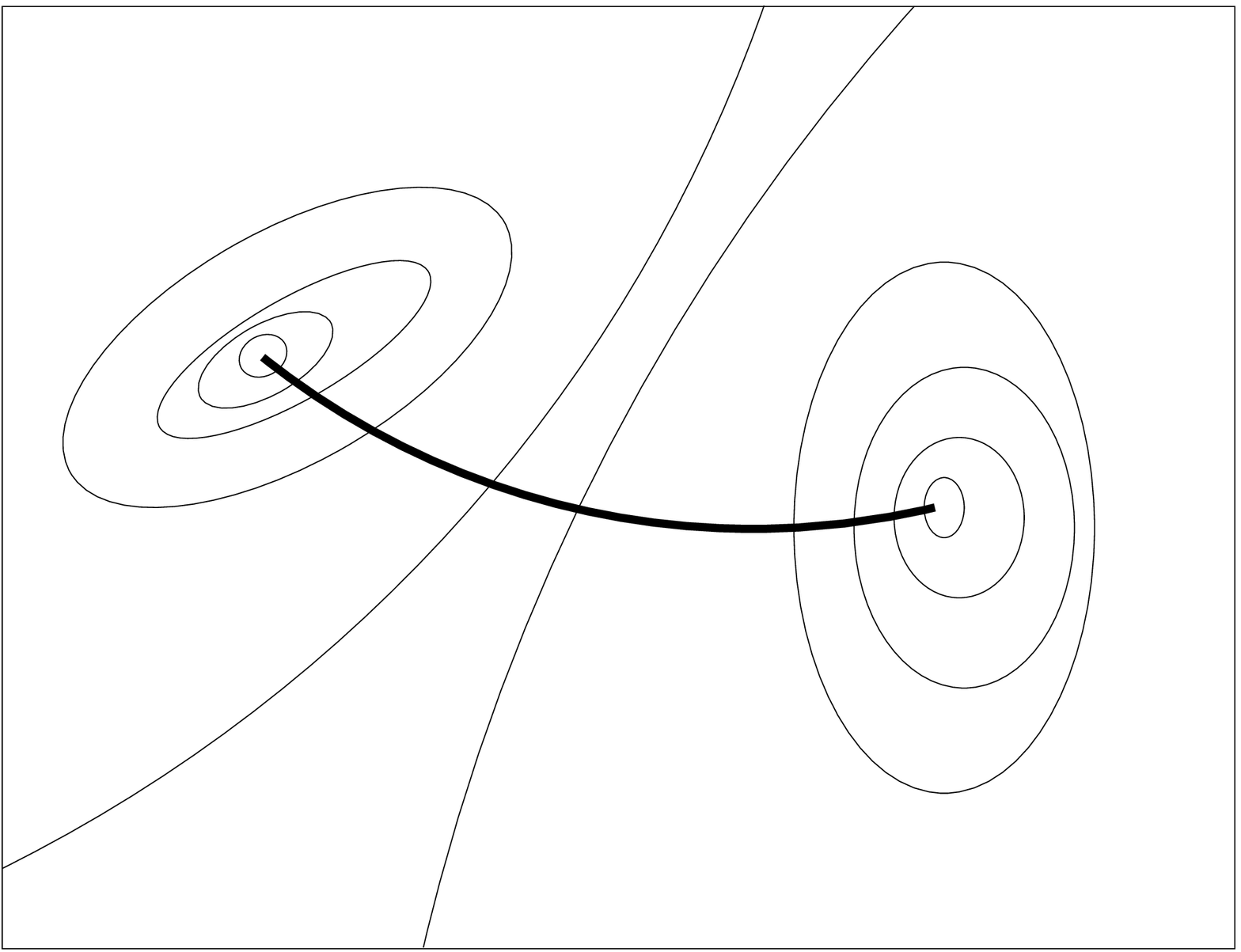}
\begin{center}
(c)
\end{center}
\end{minipage}
\begin{minipage}[t]{7cm}
\epsffile{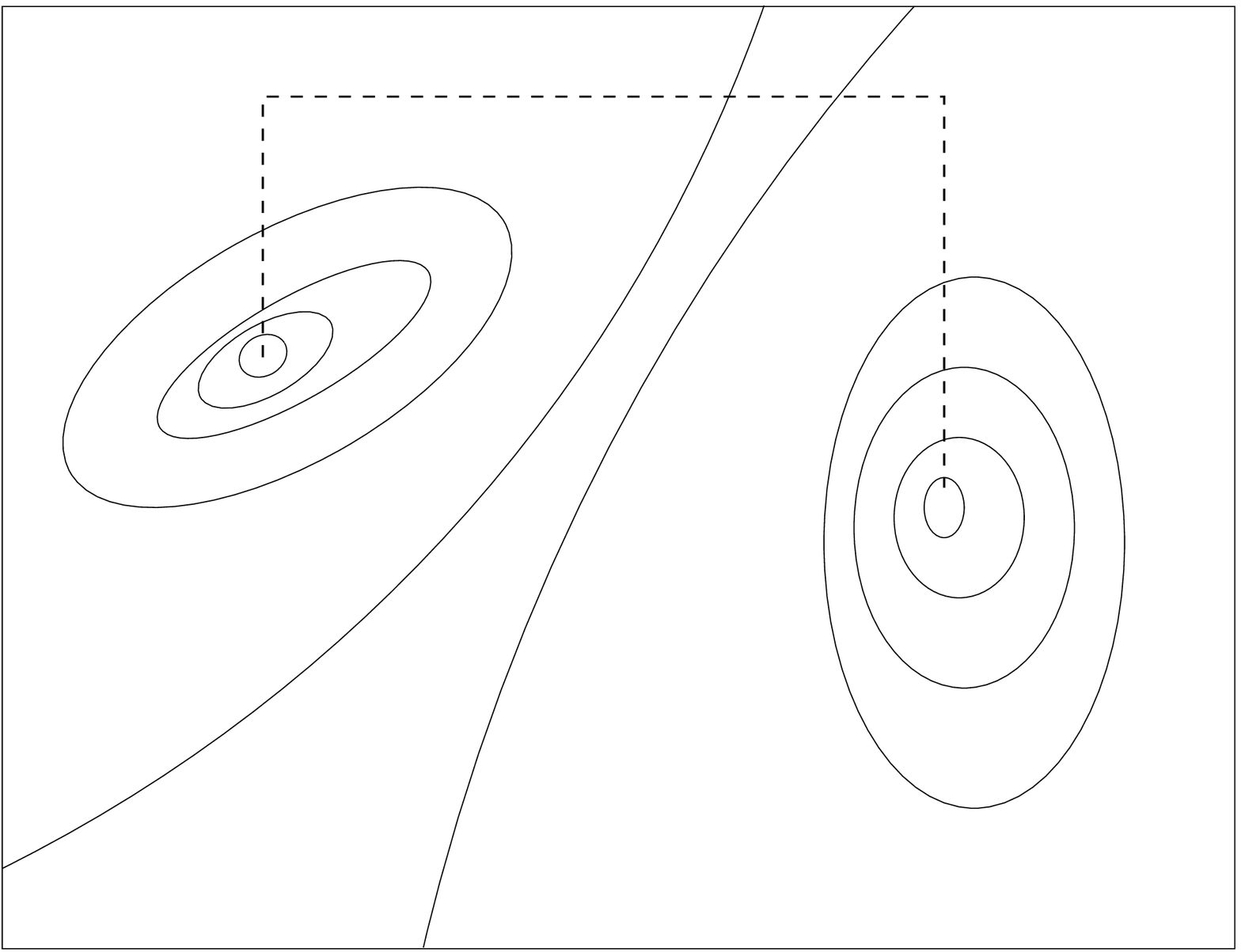}
\begin{center}
(d)
\end{center}
\end{minipage}

\vspace*{1cm}

\caption{
\label{fig:configspace}
Schematic representation of the four conceptually-different paths (the
heavy lines) one may utilize to attack the phase-coexistence problem.
Each figure depicts a configuration space spanned by two macroscopic
properties (such as energy, density \ldots ); the contours link
macrostates of equal probability, for some given
conditions $\protect\smallcon$ (such as temperature, pressure \ldots).  The
two mountain-tops locate the equilibrium macrostates associated with
the two competing phases, under these conditions.  They are separated
by a probability-ravine (free-energy-barrier). In case (a) the path
comprises two disjoint sections confined to each of the two phases and
terminating in appropriate reference macrostates. In (b) the path
skirts the ravine. 
In (c) it passes through the ravine. 
In (d) it leaps the ravine.  }

\end{figure}

It may be evident (it should certainly not be surprising) that the
extended MC framework needed to solve the phase-equilibrium problem
entails exploration of a path that {\em links} the macrostates of the
two competing phases, for the desired physical conditions $\bigcon$.  The
generic choices here are distinguished by the way in which the path is
{\em routed} in relation to the key landmark in the configuration
space  --the two-phase region which separates
the macrostates of the two phases, and which confers on them their (at
least meta-) stability. Figure~\ref{fig:configspace} depicts four
conceptually different possibilities.

First (Fig.~\ref{fig:configspace}(a)) the route may comprise two
distinct sections, neither encroaching into the two-phase region, and
each terminating in a {\em reference} macrostate. By {\em reference}
macrostate we mean one whose partition function (and thus free energy)
is already known -- on the basis of exact calculation or previous
measurement.  The information accessible through MC study of the two
sections of such a path has to be combined with the established
properties of the reference macrostates to provide the desired link
between the two equilibrium macrostates of interest. This is the traditional strategy for addressing the
phase-coexistence problem.

In the case of a liquid-gas phase boundary it is possible to choose a
path (Fig.~\ref{fig:configspace}(b)) that links the macrostates of the
two phases {\em without} passing through the two-phase region: one
simply has to sail around the critical point at which the
phase-boundary terminates. The dependence on the existence of an
adjacent critical point limits the applicability of this strategy.

The next option (it seems to be the only one left after (a) and (b); but see 
(d)) is a route which negotiates the probability ravine  
(free-energy barrier) presented by the two-phase 
region (Fig.~\ref{fig:configspace}(c)). The extended sampling tools
we discuss in the next section are essential here; with their refinement in 
recent years this route has become increasingly attractive.

If either of the phases involved is a {\em solid}, there are {\em
additional} reasons (we shall discuss them later) to avoid the ravine,
over and above the low canonical probability of the macrostates that
lie there. It is possible to do so. The necessary strategy (recently
developed) is depicted in Fig.~\ref{fig:configspace}(d). As in (a) the
path comprises two segments, each of which lies within a single phase
region. But in contrast to (a) the special macrostates to which these
segments lead are not of the traditional reference form: the defining
characteristic of {\em these} macrostates is that they they should act
as the ends of a `wormhole' along which the system may be transported
by a collective Monte Carlo move, from one to phase to the other. In
this case extended sampling methods are used to locate the wormhole
ends.

\subsection{Generic sampling strategies \label{subsec:pathsampling}}

The task of exploring (sampling) the macrostates which form a path can
be accomplished in a number of conceptually different ways; we
identify them here in a rudimentary way. We defer to subsequent sections the
discussion of specific examples which will show how the information they 
provide is used to give the result we seek; and their strengths and weaknesses.

\subsubsection*{Serial sampling \label{subsubsec:serialsampling}}

The most obvious way of gathering information about the set of macrostates
$\bigconset$ is to use canonical Boltzmann sampling to explore them one at a 
time, with a sampling distribution set to 
\begin{equation}
\label{eq:serialsampling}
\psamp(\qset) = \pcanon(\qset|\bigcon_j) \hspace{0.2cm} (j=1\ldots \Omega)
\end{equation}
in turn. One must then combine the information generated in  these $\Omega$
independent simulations. The traditional approach to our problem
(`integration methods', Section~\ref{subsec:integration}) 
employs this strategy; its basic merit is that it is simple.

\subsubsection*{Parallel sampling \label{subsubsec:parallelsampling}}

Instead of exploring the path-macrostates serially we may choose to
explore them in parallel. In this framework we simulate a set of
$\Omega$ replicas of the physical system, with the $j$-th member
chosen to realize the conditions $\bigcon_j$. The sampling distribution
in the composite configuration space spanned by $\paraqset$
is
\begin{equation}
\label{eq:parallelsampling}
\psamp(\paraqset) = \prod_{j=1}^{\Omega} \pcanon(\qset\,^{(j)}|\bigcon_j) 
\end{equation}
This is not simply a way of exploiting the availability of parallel 
computing architectures to treat $\Omega$ tasks at the same time; more significantly it  provides a way
of breaking out of the straight-jacket of local-update algorithms. The composite ensemble can be updated through interchanges of the 
coordinate sets associated with adjacent macrostates ($j$ and $j+1$ say)
to give updated coordinates
\begin{equation}
\label{eq:parallelupdate}
\hspace*{0.5cm}
\qsetprime\,^{(j)} =\qset\,^{(j+1)}
\hspace*{0.5cm}
\mbox{and}
\hspace*{0.5cm}
\qsetprime\,^{(j+1)} =  \qset\,^{(j)}
\end{equation}
The chosen sampling distribution (Eq.~\ref{eq:parallelsampling}) is
realized if such configuration-exchanges are accepted with the
appropriate probability (Eqs.~\ref{eq:metropacc},
~\ref{eq:metropaccsimple}) reflecting the change incurred in the total
energy of the composite system. This change is nominally `macroscopic'
(scales with the size of the system) and so the acceptance probability will
remain workably large only if the parameters of adjacent macrostates
are chosen sufficiently close to one another. In practice this means 
utilizing a number $\Omega$ of  macrostates that is proportional to 
$\sqrt{N}$ \cite{iba}.

Interchanging the configurations of adjacent replicas
is one instance (see  Section~\ref{subsec:switch} for another) of a {\em global-update} in which
all the coordinates evolve simultaneously. The pay-off, potentially,
is a more rapid evolution around coordinate space --more formally a
stronger mixing of the Markov chain. This is of course a general
desideratum of any MC framework. Thus it is not surprising that
algorithms of this kind have been independently devised in a wide
variety of disciplines, applied in a correspondingly wide set of
contexts \ldots and given a whole set of different names \cite{iba}. These
include Metropolis Coupled Chain \cite{geyerthomson}, Exchange Monte
Carlo \cite{hukushima}, and Parallel Tempering
\cite{marinariACS}. In Section~\ref{subsec:paracp} we shall see how one 
variant has been applied to deal with the phase-coexistence problem in 
systems with a critical point.

\subsubsection*{Extended sampling \label{subsubsec:extendedsampling}}
We will use the term {\em extended sampling} (ES) to refer to an algorithm
that allows exploration of a region of configuration space which is
`extended' with respect to the range spanned by canonical Boltzmann
sampling --specifically, one which assembles the statistical
properties of our set of macrostates $\bigcon_1 \ldots \bigcon_\Omega$
within a {\em single} simulation. Again it is straightforward to write
down the generic form of a sampling distribution that will achieve
this end; we need only a `superposition' of the canonical sampling
distributions for the set of macrostates. 
\begin{equation}
\label{eq:extendedsamplinga}
\psamp(\qset) = W_0\sum_{j=1}^{\Omega} \pcanon(\qset|\bigcon_j)
\end{equation}
where $W_0$ is a normalization constant. The superficial similarity 
between this form and those prescribed in  
Eqs~\ref{eq:serialsampling} and ~\ref{eq:parallelsampling} camouflages a 
crucial difference. Each of the distributions $\pcanon(\qset|\bigcon_j)$ 
involves a normalization constant identifiable as (Eqs.~\ref{eq:gendistdef},
\ref{eq:partialpfdef})
\begin{equation}
\label{eq:normconstants}
\weight _j  = \genpf{}{\bigcon_j}^{-1}
\end{equation} 
We do not need to know the set of normalization constants $\weightset$ to 
implement {\em serial} sampling (Eq.~\ref{eq:serialsampling}) since 
each features {\em only} as the normalization constant for a {\em sampling} 
distribution, which the MC framework does not require. Nor do we need these 
constants in implementing {\em parallel} sampling 
(Eq.~\ref{eq:parallelsampling}) since in this case they feature 
(through their product) only in the one overall normalization constant for the 
sampling distribution. But Eq.~\ref{eq:extendedsamplinga} is different. 
Writing it out more explicitly
\begin{equation}
\label{eq:extendedsamplingb}
\psamp(\qset) = W_0\sum_{j=1}^{\Omega} w_j e^{-\gene{\qset}{\bigcon_j}}
\end{equation}
we see that the weights $\weightset$ control the relative
contributions which the macrostates make to the sampling
distribution. While we are in principle at liberty to choose whatever
mixture we please (we do not {\em have} to make the assignment prescribed by Eq.~\ref{eq:normconstants})
it should be clear intuitively (we shall develop this point
in section~\ref{subsubsec:traversestrategy}) 
that the choice should confer {\em roughly} equal probabilities 
on each macrostate, so that
all are well-sampled.  It is not hard to see that the
weight-assignment made in Eq.~\ref{eq:normconstants} is in fact what we need
to fulfil this requirement. Evidently, to make {\em extended} sampling
work we {\em do} need to `know' the weights $w_j= \genpf{}{\bigcon_j}^{-1}$.
There is an element of circularity here which needs to be recognized. Our prime
objective is to determine the (relative) configurational weights of
{\em two} macrostates (those associated with two different phases, under
the same physical conditions \cite{moreexplicitly}); 
to do so (somehow or other --we
haven't yet said how) by extended sampling requires knowledge of the
configurational weights of a {\em whole path's-worth} of macrostates.
There {\em is} progress here nevertheless. 
While the two macrostates of interest are 
remote from one another, the path (by construction) comprises macrostates 
which are contiguous; it is relatively easy to determine the relative 
weights of pairs of contiguous macrostates, and thence the relative weights 
of all in the set. In effect the extended sampling 
framework allows us to replace one hard problem with a large number of 
somewhat easier problems. 

The machinery needed to turn this general strategy into a practical method 
(`building the extended sampling distribution' or `determining the macrostate 
weights') has evolved over the years from a process of trial and error to algorithms that are systematic 
and to some extent self-monitoring. The workings of the machinery is more interesting
than it might sound; we will discuss some aspects of what is involved
in Sections~\ref{subsec:traverse} and ~\ref{subsec:switch}. But we 
relegate more technical discussion (focused on recent advances) to  
Appendix~\ref{APP:WEIGHTS}. Here we continue with a broader brush.

It would be hard to write a definitive account of the development of
extended sampling methods; we will not attempt to do so. The seminal
ideas are probably correctly attributed to Torrie and Valleau
\cite{torrvalleau} who coined the terminology {\em umbrella sampling}.
The huge literature of subsequent advances and
rediscoveries may be rationalized a little by dividing it into two, according 
to how the macrostates to be weighted are defined.

If the macrostates are defined by a set of values $\lambdavalset$
of some generalized `field' $\lambda$, the sampling distribution is of 
the form   
\begin{equation}
\label{eq:extendedsamplingfield}
\psamp(\qset) = W_0\sum_{j=1}^{\Omega} w_j e^{-\genesimp (\qset, \lambda_j)}
\end{equation}
Extended sampling strategies utilizing this kind of representation feature in 
the literature with a variety of titles: {\em expanded ensemble} \cite{lyub},
{\em simulated tempering} \cite{marinariparisi}, {\em temperature scaling}
\cite{valleau}.

On the other hand, if the macrostates are defined on
some `macrovariable' $\plainop$ the sampling distribution is of the form   
\begin{equation}
\label{eq:extendedsamplingdensity}
\psamp(\qset) = W_0\sum_{j=1}^{\Omega} w_j e^{-\genesimp (\qset)} 
\deltapick{j}{\plainop}
\end{equation}
where
\begin{equation}
\label{eq:pickmacro}
\deltapick{j}{\plainop}
\equiv 
\left\{ 
\begin{array}{ll}
1 & {\plainop} \in \mbox{range associated with } {\bigcon _j} \\
0 & \mbox{otherwise}
\end{array}
\right.
\end{equation}
Realizations of this formalism go under the names {\em adaptive
umbrella sampling} \cite{mezei} and the {\em multicanonical ensemble}
introduced by Berg and co-workers \cite{bn}. It seems right to
attribute the recent revival in interest in extended sampling to the
latter work. 

In Sections~\ref{subsec:traverse} and
\ref{subsec:switch} we shall see that extended sampling strategies provide a rich variety of 
ways of tackling the phase coexistence problem, including the distinctive 
problems arising when one of the phases is of solid form.

\section{Path-based techniques: a guided tour \label{SEC:PATHBASEDTECHNIQUES}}

We now proceed to explore how the strategic options in regard to
routing and sampling of paths (Sections~\ref{subsec:pathrouting} and
\ref{subsec:pathsampling}) can be fused together to form practical 
techniques for addressing the 
phase-coexistence problem. We do not set out to be exhaustive here:
there are very many pairings of routes and sampling strategies (the
choices are to some extent mutually independent, which is why we
discussed them separately). We focus rather on a few key cases; we
explain what information is gathered by the chosen sampling of the
chosen path and how it is used to yield the desired free-energy
comparison; and we assess the strengths and weaknesses of each
technique as we go.

\subsection{Keeping it simple: numerical integration and 
reference states \label{subsec:integration}}

\subsubsection{The strategy \label{subsubsec:integrationstrategy}}

The staple approach to the phase-coexistence problem involves serial
sampling (Section \ref{subsubsec:serialsampling}) along a path of the type
depicted in Figure~\ref{fig:configspace}(a). Effectively the single
core problem (comparing the configurational weights --free-energies--
of the two physical macrostates) is split into two, each requiring
comparison of a physical macrostate ($\smallcon, \alpha$) with some suitable
reference macrostate, $\bigconref_{\alpha}$. A reference macrostate will be 
`suitable' if one can identify a path parameterized by some field
$\lambda$ linking it to the physical macrostate, with 
$\lambda = \lambda_1$ and $\lambda =\lambda_{\Omega}$ 
denoting respectively the physical and the
reference macrostates.  The sampling distribution 
at some arbitrary point, $\lambda$, on this path is then of the form
(Eqs.~\ref{eq:partialdistdef},\ref{eq:partialpfdef}, \ref{eq:serialsampling})
\begin{equation}
\label{eq:integrationone}
\psamp(\qset) = \pcanon(\qset|\lambda) =
\frac{1}{\genpf{\alpha}{\lambda}} e^{-\gene{\qset}{\lambda}}
\end{equation}
with
\begin{equation}
\label{eq:integrationtwo}
\genpf{\alpha}{\lambda } 
\equiv 
\intdq \deltapick{\alpha}{\plainop (\qset)}  e^{-\gene{\qset}{\lambda}} 
\equiv 
e^{-\genf{\alpha}{\lambda}}
\end{equation}

We have seen that free-energies like $\genf{\alpha}{\lambda}$ are not
themselves naturally expressible as canonical averages; but their
{\em derivatives} with respect to field-parameters {\em are} expressible
this way. Specifically,
\begin{eqnarray}
\frac
{\partial \genf{\alpha}{\lambda}}
{\partial \lambda} 
&=& 
-\frac{1}{\genpf{\alpha}{\lambda}}
\frac{\partial \genpf{\alpha}{\lambda}}{\partial  \lambda}
\nonumber\\
&=& \frac{1}{\genpf{\alpha}{\lambda}}
\intdq \deltapick{\alpha}{\plainop (\qset)} 
\frac{\partial \gene{\qset}{\lambda}}{\partial\lambda}
 e^{-\gene{\qset}{\lambda}} 
\nonumber\\
&=& 
\expec{\alpha, \lambda}{\frac{\partial \gene{\qset}{\lambda}}{\partial\lambda}}
\label{eq:integrationthree}
\end{eqnarray}
where the average is to be taken with respect to the canonical 
distribution for macrostate $\alpha, \lambda$ (Eq.~\ref{eq:integrationone}.)
The free-energy difference between the physical and reference macrostates
is thus formally given by 
\begin{equation}
\genf{}{\bigconref_\alpha}
-\genf{\alpha}{\smallcon}
=
\int_{\lambda_1}^{\lambda_{\Omega}}d\lambda \frac
{\partial \genf{\alpha}{\lambda}}
{\partial \lambda} 
= 
\int_{\lambda_1}^{\lambda_{\Omega}}d\lambda 
\expec{\alpha, \lambda}{\frac{\partial \gene{\qset}{\lambda}}{\partial\lambda}}
\label{eq:lambdaint}
\end{equation}
A sequence of independent simulations, conducted at a set of points 
$\lambdavalset$ spanning a path from $\lambda_1$ to $\lambda_\Omega$, 
then allows one to
estimate the free energy difference by numerical quadrature:
\begin{equation}
\label{eq:lambdasum}
\genf{}{\bigconref_\alpha}
-\genf{\alpha}{\smallcon}
\stackrel{eb}{=} 
\sum_{j=1}^{\Omega} \expec{\alpha ,\lambda_j}{\frac{\partial \gene{\qset}{\lambda}}{\partial\lambda_j}} \Delta \lambda
\end{equation}

This takes us half way. The entire procedure has to be repeated for the second 
phase $\alphatilde$,
integrating along some path parameterized (in general) by some 
{\em other} field 
$\tilde{\lambda}$ running 
between the macrostate 
$\alphatilde, \smallcon$ and some {\em other} reference state 
$\bigconref_{\alphatilde}$. Finally the results of the two procedures are 
combined to give the quantity of interest (Eq.~\ref{eq:deltaf})
\begin{equation}
\label{eq:deltafintegration}
\Delta \genf{\alphaalphatilde}{\smallcon}
= 
\genf{}{\bigconref_{\alpha}}
-\genf{}{\bigconref_{\alphatilde}}
- 
\Delta \lambda\sum_{j=1}^{\Omega} \expec{\alpha,\lambda_j}{\frac{\partial \gene{\qset}{\lambda}}{\partial\lambda_j}} 
+
\Delta \tilde{\lambda}\sum_{j=1}^{\tilde{\Omega}} \expec{\alphatilde, \tilde{\lambda}_j}{\frac{\partial \gene{\qset}{\tilde{\lambda}}}{\partial \tilde{\lambda}_j}} 
\end{equation}

We shall refer to this strategy as {\em numerical integration to reference 
macrostates} (NIRM). It is helpful to assess its strengths and weaknesses armed 
with an explicit example. 

We shall consider what is arguably the archetypal example of the NIRM
strategy: the Einstein Solid method (ESM) \cite{frenkelladd}.
The ESM provides a
simple way of computing the free energies of crystalline phases, and
thence addressing questions of the relative stability of competing
crystalline structures.  We describe its implementation for the
simplest case where the inter-particle interaction is of hard-sphere
form; it is readily extended to deal with particles interacting
through soft potentials \cite{FRENKELSMIT}.
 
The name of the method reflects the choice of reference macrostate: a
crystalline solid comprising particles which do not interact with one
another, but which are bound by harmonic springs
to the sites of a crystalline lattice, 
$\Rset{\alpha}$, coinciding with 
that of the phase of interest.

The relevant path is constructed from sampling distributions of the general 
form prescribed in Eq.~\ref{eq:integrationone} with
\begin{equation}
\label{eq:ESMone}
\gene{\qset}{\lambda} = \gene{\rset}{\smallcon} + 
\lambda\sum_{i=1}^N \left( \vec{r}_i-\vec{R}^{\alpha}_i \right)^2 
\end{equation}
The first term on the RHS contains the hard-sphere interactions.
The second term embodies the harmonic spring
energy, which reflects the displacements of the particles from their
lattice sites. In this case the point $\lambda_1=0$ locates the
physical macrostate.
With increasing $\lambda$, the energy cost associated
with a given set of displacements increases, with a concomitant reduction in
the size of the typical displacements. On further increasing $\lambda$,
one ultimately reaches a point $\lambda_\Omega$ beyond which particles
are so tightly bound to their lattice sites they practically never
collide with one another; the hard sphere interaction term then plays
no role, thus realizing the desired reference macrostate, whose
free-energy may be computed exactly. 

Some of the results of ESM studies of crystalline phases of hard spheres
are shown in Table~\ref{table:hardspheredata}. We shall discuss them below.

\subsubsection{Critique \label{subsubsec:integrationcritique}}

There is much to commend NIRM: it is conceptually simple; it can be
implemented with only a modest extension of the simulation framework
already needed for standard MC sampling; and it is versatile.  It has
been applied successfully in free-energy measurements of ({\it inter alia})
crystalline solids \cite{frenkelladd}, liquids \cite{INTliquids} and
liquid crystals \cite{INTliquidcrystals}. It is probably regarded as
{\em the} standard method for attacking the phase-coexistence problem,
and it provides the benchmark against which other approaches must be
assessed.  Nevertheless (as one might guess from the persistence of
attempts to develop `other approaches') it is less than ideal in a
number of respects. We discuss them in turn.

First, the NIRM method hinges on the {\em identification of a good
path and reference macrostate.}  A `good' path is short; but the reference
macrostate (the choice of which is limited) may lie far from the physical
macrostate of interest, entailing a large number of independent
simulations to make the necessary link. In a sense the ESM provides a
case in point: the reference macrostate is strictly located at
$\lambda_{\Omega} =\infty$; corrections for the use of a finite
$\lambda_{\Omega}$ need to be made \cite{frenkelladd}.  This kind of
problem is a nuisance, but no worse. A potentially more serious
constraint on the path is that the derivative being measured should
vary slowly, smoothly and reversibly along it; if it does not the
numerical quadrature may be compromised. A phase transition en route
(whether in the `real' space of the physical system or the
extended-model-space into which NIRM simulations frequently extend) is
thus a particular hazard. The realization of NIRM known as the Single
Occupancy Cell method (SOCM) \cite{HR68} provides an example where
such concerns arise, and seem not to have been wholly dispelled
\cite{ogura}.

The {\em choice of simulation parameters} also raises issues.
Evidently one has to decide how many simulations are to be performed
along the path and at which values of $\lambda$. In so doing one must
strike a suitable balance between minimizing computation time while
still ensuring that no region of the path (particularly one in which
the integrand varies strongly) is neglected. This may necessitate
a degree of trial and error.

The {\em uncertainties} to be attached to NIRM estimates are
problematic in a number of respects. Use of simple numerical
quadrature (estimating the integral in Eq.~\ref{eq:lambdaint} by the
sum in Eq.~\ref{eq:lambdasum}) will result in errors. One can reduce
such errors by interpolation into the regions of $\lambda$ between the
chosen simulation values (for example using histogram re-weighting
techniques discussed in Appendix~\ref{APP:HISTOGRAMREWEIGHTING});
but there will still be systematic errors associated both with the
interpolation and with finite sampling times. No reliable and
comprehensive prescription for estimating the magnitude of such errors
has yet been developed.

These problems are exacerbated when one addresses the quantity of real
interest --the {\em difference} between the free energies of two
competing phases. The fact that NIRM treats this as {\em two} problems
rather than {\em one} (Eq.~\ref{eq:deltafintegration}) is problematic in two
respects.  

First this aspect of the NIRM strategy {\em compounds} a problem
inherent in all simulation studies of this problem (or, indeed, any
other in many-body physics): finite-size effects.  This issue deserves
a section to itself, and gets 
one (Section~\ref{subsec:finitesize}). Here we note simply that such
effects are harder to assess when one has to synthesize calculations
on different phases, utilizing different reference states, and
different system sizes --and sometimes conducted by different authors.

Second, since the entire enterprise is constructed so as to locate
points (of phase equilibrium) at which the free-energy 
difference {\em vanishes}, in NIRM
one is inevitably faced with the task of determining some very small
number by taking the difference between two relatively large
numbers. This point is made more explicitly by the hard-sphere data in
Table~\ref{table:hardspheredata}.
One sees that the {\em difference}
between the values of the free energy \cite{purelyentropic} 
of the two
crystalline  phases is some {\em four orders of
magnitude} smaller than the separate results for the two phases,
determined by ESM. Of course one can see this as a testimony to the
remarkable care with which the most recent recent ESM studies have
been carried out \cite{pronkfrenkelone}. Alternatively one may see it
as a strong indicator that another approach is called for. 

\subsection{Parallel tempering; around the critical point
\label{subsec:paracp}}

\subsubsection{The strategy \label{subsubsec:paracpstrategy}}

When the coexistence line of interest terminates in a critical point
the two phases can be linked by a single continuous path
(Figure~\ref{fig:configspace}(b)) which loops around the critical
point, eliminating the need for reference macrostates, while still avoiding 
the inter-phase region. In principle it is possible to establish the
location of such a coexistence curve by integration along this route. But the techniques of parallel sampling 
(Section~\ref{subsubsec:parallelsampling}) provide a substantially
more elegant way of exploiting such a path, in a technique known as (hyper) 
parallel tempering (HPT) \cite{depablo}.

\begin{figure}
\setlength{\epsfxsize}{14.0cm}
\centerline{\mbox{\epsffile{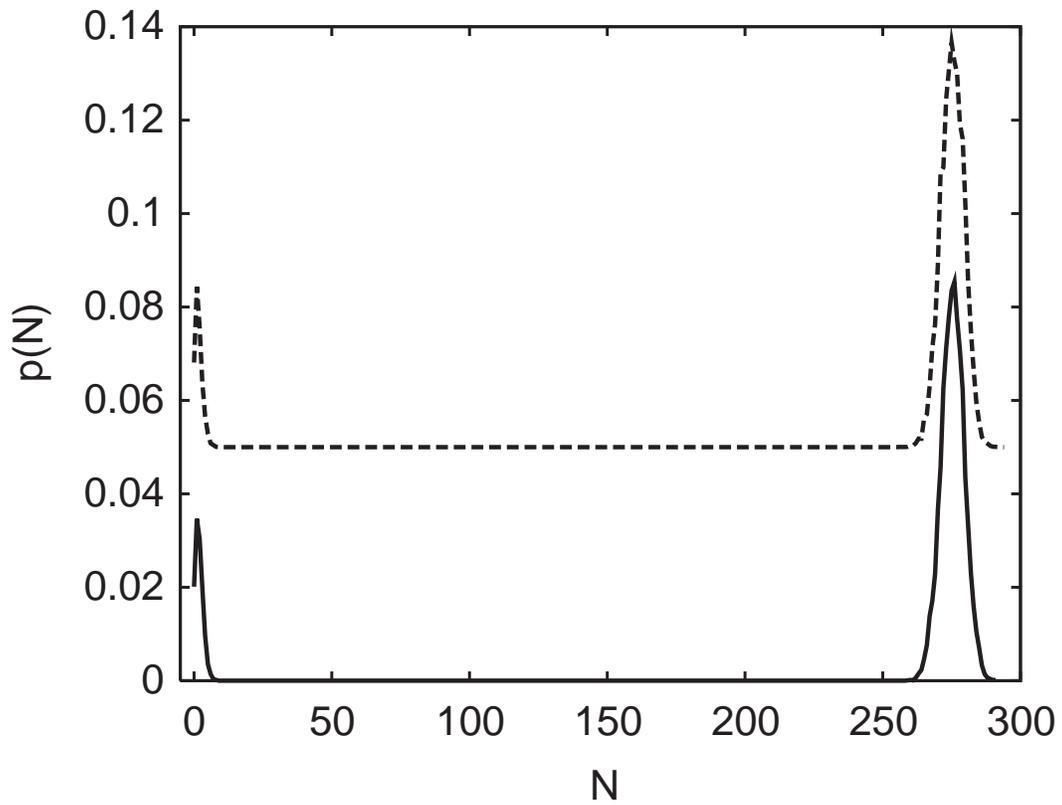}}}

\vspace*{1cm}
\caption{
Probability distribution of the number $N$ of particles in a LJ fluid
The simulations use the HPT method 
described in section ~\ref{subsubsec:paracpstrategy}. 
The solid line shows the distribution
for a replica whose $\mu - T$ parameters lie close to coexistence;
the dashed line (offset) shows the distribution (for the same 
$\mu -T$ parameters)  obtained by folding in (explicitly) the 
contributions of all replicas, using
multi-histogram re-weighting.  
\protect\figcite{2}{depablo}
}

\label{fig:depabloone}
\end{figure}

\begin{figure}
\setlength{\epsfxsize}{14.0cm}
\centerline{\mbox{\epsffile{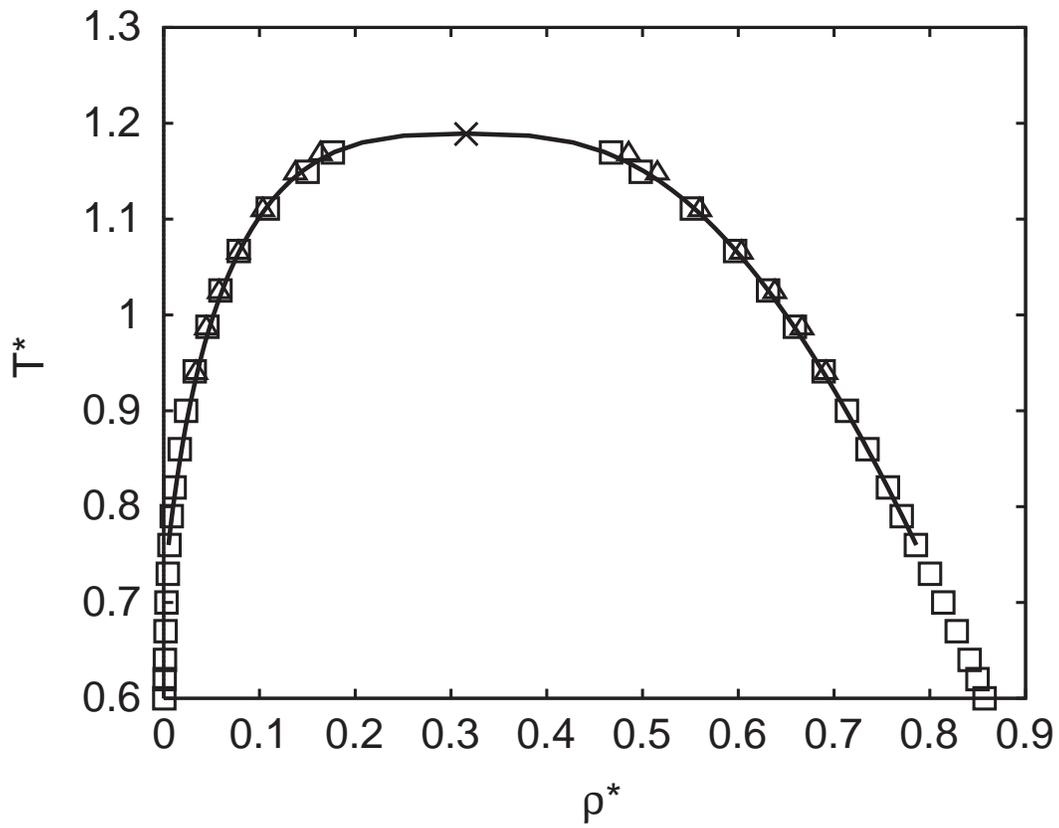}}} 

\vspace*{1cm}

\caption{ Phase diagram for the LJ fluid. The squares show results 
obtained by the HPT method
described in section ~\ref{subsubsec:paracpstrategy}
and illustrated in Figure~\ref{fig:depabloone}. 
The triangles show results obtained by the ESIT strategy
described in section ~\ref{subsubsec:traversestrategy}
and illustrated in figure~\ref{fig:nbwlj}.
\protect\figcite{3}{depablo}
}

\label{fig:depablotwo}
\end{figure}

Studies of liquid-vapor coexistence are, generally, best addressed in
the framework of an open ensemble; thus the state variables here
comprise both the particle coordinates $\rset$ and the particle
number $N$. A path with the appropriate credentials can be constructed by 
identifying pairs of values of the chemical potential $\mu$ and the 
temperature $T$ which trace out some rough {\em approximation} to the
coexistence curve in the $\mu-T$ plane, but extend 
into the one-phase region beyond the critical point.
Once again there is some circularity here to which we shall return.
Making the relevant variables explicit,
the sampling distribution (Eq.~\ref{eq:parallelsampling}) takes the form
\begin{equation}
\label{eq:paracpone}
\psamp
(
\rset^{(1)},N^{(1)} 
\ldots
\rset^{(\Omega)},N^{(\Omega)}
) 
= \prod_{j=1}^{\Omega} \pcanon(\rset^{(j)}, N^{(j)}|\mu_j, T_j) 
\end{equation}
In the context of liquid-vapor coexistence the particle number $N$
(or equivalently the number density $\rho=N/V$) plays the role of an 
order parameter.
Estimates of the 
distribution $\pcanon(N|\mu, T)$ are available from the simulation for
all the points chosen to define the path. One may then identify the 
free energy difference from the integrated areas of the branches of 
this distribution associated with each phase and proceed to search for coexistence using the
criterion that these integrated areas should be equal 
(Eq.~\ref{eq:deltaf} et seq.). 
Figures~\ref{fig:depabloone}
and \ref{fig:depablotwo} show some explicit results \cite{depablo} for a Lennard-Jones fluid.

\subsubsection{Critique \label{subsubsec:paracpcritique}}

The phase-diagram shown in Figure~\ref{fig:depablotwo} is of course
the ultimate objective of such studies; but it is the {\em
distribution} shown in Figure~\ref{fig:depabloone} that merits most
immediate comment, lest its key feature be taken for-granted.  Its `key
feature' (which we shall come to recognize as the signature of
success in this enterprise) is that it captures the contributions
from {\em both} phases (the two distinct peaks) within the framework
of a {\em single simulation}. The fact that both phases are
well-visited shows that this strategy manages to break the ergodic
block which dooms conventional MC sampling to explore only the
particular phase in which the simulation happens to be launched
(Section~\ref{subsec:tools}).

The HPT method deals with the ergodic block by avoiding it. The
configuration exchange between adjacent replicas
(Eq.~\ref{eq:parallelupdate}) fuels a form of continuous tempering: a
liquid phase configuration resident in a replica low down the
coexistence curve may diffuse `along the path' and thus through the
replicas, to a point (perhaps super-critical) at which on-going local
updating is effective in eroding memory of its liquid origins; that (ever evolving) configuration may then diffuse downwards, to appear in the
original replica as a vapor phase configuration.

\begin{figure}
\setlength{\epsfxsize}{14.0cm}
\centerline{\mbox{\epsffile{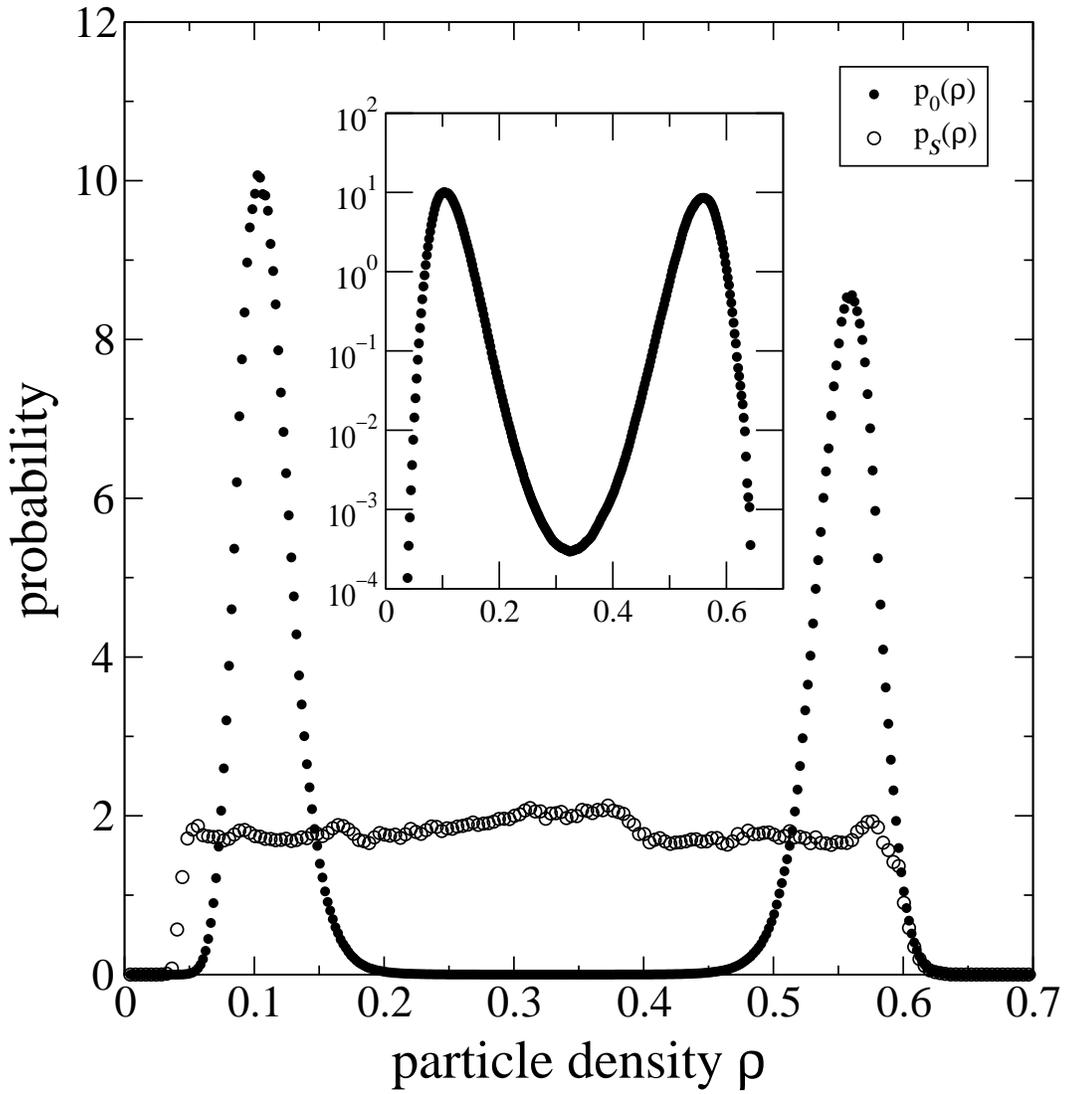}}}

\vspace*{1cm}

\caption{Results from a multicanonical simulation of the 3D
Lennard-Jones fluid at a point on the coexistence curve. 
The figure shows both the multicanonical sampling 
distribution $\psamp(\rho)$ (symbols: $\circ$) 
and the corresponding estimate of
the equilibrium distribution $P_{\equm}(\rho)$ with 
$\rho =N/V$ the number density. The inset shows the value 
of the  equilibrium distribution in the interfacial region. 
\protect\cite{nbwljunpub}.
}
\label{fig:nbwlj}
\end{figure}

There is an elegant and powerful idea here, albeit one whose
applicability (to the phase-coexistence problem) is limited to systems
with critical points. But there is one respect in which it is less
than satisfactory --the element of circularity already noted: the
method will tell us {\em if} we have selected a point sufficiently
close to coexistence for both phases to have observable probability;
it does not in itself tell us what to do if our selection does not
satisfy this criterion. How `close' we need to be depends sensitively
on the size of the system simulated.  The ratio of the two phase
probabilities (at a chosen point in field-space) varies exponentially
fast with the system size \cite{variationwithN}.  Thus for a `large'
system, unless we are `very close' to coexistence, the equilibrium
(grand-canonical) PDF determined in HPT will show signs of only one
phase {{\em even when the there is no ergodic block preventing access
to the other phase}.  If the initial choice is close enough to
coexistence to provide at least {\em some} signature of the
sub-dominant phase \cite{morerestrictivethanthis} then histogram
re-weighting techniques (Appendix~\ref{APP:HISTOGRAMREWEIGHTING}) can
be used to give a better estimate of coexistence. But HPT provides no
way of systematically improving bad initial estimates, because it
provides no mechanism for dealing with the huge difference between the
statistical weights of the two phases away from the immediate vicinity
of coexistence. To address {\em that} issue one must turn to extended
sampling techniques.

\subsection{Extended sampling: traversing the barrier \label{subsec:traverse}}

\subsubsection{The strategy \label{subsubsec:traversestrategy}}

Viewed from the perspectives of configuration space provided by the
caricature in Figure~\ref{fig:configspace} the most direct approach
to the phase-coexistence problem calls for a full frontal assault on
the ergodic barrier that separates the two phases. The extended sampling 
strategies discussed in Section~\ref{subsubsec:extendedsampling} make that 
possible. The framework we need is  a synthesis of Eqs.~\ref{eq:deltaf} and 
\ref{eq:extendedsamplingdensity}. We will refer to it generically as 
Extended Sampling Interface Traverse (ESIT).

Equation~\ref{eq:deltaf} shows that we can always accomplish our
objective if we can measure the full canonical distribution of an
appropriate order parameter. By `full' we mean that the contributions
of both phases must be established {\em and calibrated on the same
scale}. Of course it is the last bit that is the problem. (It is always
straightforward to determine the two {\em separately} normalized
distributions associated with the two phases, by conventional sampling
in each phase in turn.) The reason that it is a problem is that the
`full canonical' distribution of the (an) `order parameter' is
typically vanishingly small at values intermediate between those
characteristic of the two individual phases. 
The vanishingly small values provide
a real, even quantitative, measure of the ergodic barrier between the
phases. If the `full' order parameter distribution is to be determined by a
`direct' approach (as distinct from the circuitous approach of
Section~\ref{subsec:paracp}, or the `off the map' approach to be discussed in
Section~\ref{subsec:switch}) these low-probability macrostates {\em must} be
visited.

Equation~\ref{eq:extendedsamplingdensity} shows how. We need to build
a sampling distribution that `extends' along the path of
$\plainop$-macrostates running between the two phases. To do its job
that sampling distribution must
(Section~\ref{subsubsec:extendedsampling}) assign `roughly equal'
values to the probabilities of the different macrostates. More explicitly the
resulting measured distribution of ${\plainop}$-values 
(following \cite{bn} we shall call it multicanonical)
\begin{equation}
\label{eq:traverseone}
\psamp(\plainop_j) \equiv 
\intdq \psamp(\qset) \deltapick{j}{\plainop(\qset)}
\end{equation}
should be `roughly flat'. It needs to be `roughly flat' because the 
macrostate of lowest probability sets the size of the bottleneck 
through which inter-phase traverses must pass. It needs to be 
{\em no better} than `roughly' flat because of the way in which (ultimately) 
it is {\em used}. It is used to estimate the true canonical distribution
$\pcanon(\plainop)$. The two distributions are simply related by
\begin{equation}
\label{eq:traversetwo}
\pcanon(\plainop_j)
\eqdot
w_j^{-1}\psamp(\plainop_j) 
\end{equation}
where $\weightset$ are the multicanonical weights that define the chosen sampling distribution (Eq.~\ref{eq:extendedsamplingdensity}) and $\eqdot$ means equality to within an overall normalization constant. 
The procedure by which one uses this equation to
estimate the canonical distribution from 
the measured distribution is variously referred to as 
`unfolding the weights' or `re-weighting'; it is simply one realization of 
the identity given in Eq.~\ref{eq:identity}. The procedure eliminates any {\em explicit} 
dependence on the weights (hence the looseness of the criteria by 
which they are specified); but it leaves the desired legacy:
the relative sizes of the two branches
of the
{\em canonical} distribution are determined with a statistical quality
that reflects the number of inter-phase traverses 
in the {\em multicanonical} ensemble.

This strategy has been applied to the study of a range of coexistence
problems, initially focused on lattice models in magnetism
\cite{multicanmag} and particle physics \cite{multicanparticle}.
Figure~\ref{fig:nbwlj} 
\cite{nbwljunpub,nbwlj,seealsodepablo}
shows the results of
an application to liquid-vapor coexistence in a Lennard-Jones system
with the particle number density chosen as an order parameter.

\subsubsection{Critique \label{subsubsec:traversecritique}}

The ESIT strategy has clear advantages with respect to the NIRM and 
(to a lesser extent) HPT strategies discussed in preceding sections

While HPT {\em requires} the existence of a critical point ESIT does
not (although its existence can be usefully {\em exploited} to assist in the
task of weight generation \cite{nbwlj}). 
Moreover, in contrast to HPT, ESIT does not {\em
require} boot-strapping by a rather good guess of a point on the phase
boundary. Extended sampling methods can take huge probability
differentials in their stride (100 orders of magnitude is not
uncommon) as one can see from the low probability of the macrostates
in the inter-phase ravine in Fig.~\ref{fig:nbwlj}. Thus
ESIT allows one to measure the `free energy
difference' between two phases relatively far from the phase boundary --
in fact in any regime in which both phases are at least `metastable'
\cite{trickypoint}. 

ESIT has one further potential advantage with respect to HPT, 
which emerges once one appreciates what it
entails at a {\em microscopic} level --in particular the nature of the
intermediate macrostates that lurk in the ravine between the two peaks
of the canonical distribution. To do so one needs first to
understand a {\em general} point about the weight generation
procedures used to build the sampling distribution. These procedures
do not {\em require} explicit specification of the configurations in
the macrostates along the path; rather they {\em search} for the
configurations that dominate those macrostates. The results of the
`search' can sometimes be a little unexpected \cite{funnygatewaystates,lshstwo}.
But in the case of the liquid-vapor
inter-phase path picked out by choosing the density as order
parameter, the results of the search are clear {\em a priori}: the
configurations dominating such a path will show two
physically-distinct regions, each housing one of the two phases,
separated by an interface.
The dominance of this kind of configuration
is recognized in arguments which establish the convexity of the
free-energy in the thermodynamic limit, and the way in which that
limit is approached \cite{convexity}; this picture also allows one
to understand the depth of the probability ravine to be traversed 
\cite{howbigisit}.
It follows then that ESIT
provides incidental access to these configurations, and thus to
information about their dominant feature--the interface. Information of this 
kind has been exploited in a number of studies \cite{interfacestudies}.

The broader and more far-reaching comparison to be made here is however
between NIRM and strategies {\em like} ESIT which furnish canonical
distributions spanning two phases, such as that shown in
Figure~\ref{fig:nbwlj} \cite{notjustesit}. Here it seems to us that
ESIT wins in two respects.  First, it is rather more transparent in
regard to uncertainties (in free-energy differences).
The error bounds emerging from ESIT (and
family) represent purely statistical uncertainties associated with the
measurement of the relative weights of two distribution-peaks. In
contrast NIRM error bounds have to aggregate the uncertainties
(statistical and systematic) associated with different stages of the
integration process. Second, it seems rather more satisfying to 
read-off the result for a free energy difference directly from the likes
of Fig.~\ref{fig:nbwlj} than from a pair of numbers
established by appeal to physically-irrelevant reference states.

Nevertheless the ESIT strategy is demanding
in a number of respects.  Generating the weights is a
computationally-intensive job, which is not yet fully self-managing.
Subsequent sampling of the resulting multicanonical
distribution is a slow process: the
dynamics in $\plainop$-space is a random walk in which visits to the
macrostates of low equilibrium probability are secured only at the
expense of repeated refusal of proffered moves to the dominant
equilibrium macrostates. It helps 
(though it goes against the spirit of `one simulation') to break the space
up into sections, whose length is chosen to reflect an interplay of the 
diffusion time {\em between} macrostates and the time associated with 
relaxation {\em within} macrostates \cite{frenkelargument}.

These two reservations apply generically to ES methods; the following one
is specific to ESIT itself. There are some phase coexistence problems
in which an inter-phase path involving an {\em interface} is
computationally fraught. In particular if one of the phases is
crystalline (as in the case of melting/freezing) or if both are
crystalline (there is a potential {\em structural} phase transition)
such a traverse will involve substantial, physically slow,
restructuring --- vulnerable to further ergodic traps, and compounding
the intrinsic slowness of the multicanonical sampling process 
\cite{butitdoesnthaveto,grsadbspt}.
In such circumstances it would be better if the inter-phase trip could be
accomplished without encountering interfaces. This is possible.

\subsection{From paths to wormholes: phase switch 
\label{subsec:switch}
}
We shall devote rather more time to this fourth and final example of
path-based strategies. We do so for two reasons. First it
provides us with an opportunity to touch on a variety of other
strands of thought about the `free-energy-estimation problem' which
should feature somewhere in this article, and can do so helpfully
here. And second, we like it.

The strategy we shall discuss is a way of realizing the direct leap
between phases represented in Fig.~\ref{fig:configspace}(d). In the
literature we have referred to it as `Lattice Switch Monte Carlo'
\cite{lshsone,lshstwo,lssoftpot} in the context of phase equilibrium between crystalline 
structures,  and `Phase Switch Monte Carlo'
\cite{lsfreezing} in the context of solid-liquid coexistence. Here we
shall develop the ideas in a general form; and we shall refer to the
method as Extended Sampling Phase Switch (ESPS).

\subsubsection{Strategy \label{subsubsec:switchstrategy}}

Let us return to the core problem: the evaluation of the {\em ratio}
(Eq.~\ref{eq:deltaf}) of configurational integrals of the form
prescribed in Eq.~\ref{eq:partialpfdef}. We can make the problem look
different (and possibly make it easier) by choosing to express it in
terms of coordinates that are in some sense `matched' to each 
phase. The simplest useful possibility is provided
by an appropriate {\em linear} transformation 
\begin{equation}
\label{eq:switchone}
\qset = \qsetref{\alpha} + \taumat{\alpha} (\uset)
\end{equation}
One can think of $\qsetref{\alpha}$ as some reference point
in the configuration space of phase $\alpha$ and 
$\uset$ as a `displacement' from that point, modulo some rotation or dilation, 
prescribed by the operation $\taumat{\alpha}$. 
The single-phase partition function 
in Eq.~\ref{eq:partialpfdef} can then be written in the form
\begin{equation}
\label{eq:switchtwo}
\genpf{\alpha}{\smallcon} 
=
\det \taumat{\alpha}\intdu  e^{-\genephase{\uset}{\smallcon}{\alpha}}
\end{equation}
where $\det \taumat{\alpha}$ is the Jacobean \cite{whynojacobeansbefore} 
of the transformation (a 
configuration-independent constant, given the presumed linearity)
and
\begin{equation}
\label{eq:switchthree}
e^{-\genephase{\uset}{\smallcon}{\alpha}}
=
e^{-\gene{\qset}{\smallcon}} \deltapick{\alpha}{\qset}
\end{equation}
The form of the new energy function (`hamiltonian')
$\genephase{\uset}{\smallcon}{\alpha}$
reflects the {\em representation} chosen for
the region of configuration space relevant to the phase; because this energy function
carries a phase label, it can also be used to
absorb the constraint (Eq~\ref{eq:pickphaseqset}) that restricts the
integral to that region \cite{constraintimplicit}.

The difference between the free energies of the two phases 
(Eq.~\ref{eq:deltaf}) can now be written in the form
\begin{equation} 
\label{eq:switchfour}
\Delta \genf{\alphaalphatilde}{\smallcon}
=
\Delta \genesimp^{0}_{\alpha \tilde{\alpha}} 
- \ln \det{\smat}_{\alpha \alphatilde} 
- \ln {\cal R}_{\alpha \alphatilde} 
\end{equation}
where
\begin{equation} 
\label{eq:switchfive}
\Delta \genesimp^{0}_{\alpha \tilde{\alpha}} =
\gene{\qsetref{\alpha}}{\smallcon}
-
\gene{\qsetref{\alphatilde}}{\smallcon}
\end{equation}
is the difference between the hamiltonians
of the reference configurations
while
\begin{equation} 
\label{eq:smat}
\smat _{\alpha \alphatilde} = 
\taumat{\alpha} \times \taumat {\alphatilde} ^{-1}
\end{equation}
These two contributions to Eq.~\ref{eq:switchfour} are computationally trivial;
the computational challenge is now in the third term defined by
\begin{equation}
\label{eq:configintratio}
{\cal R}_{\alpha \alphatilde} 
= \frac
{\intdu  e^{-\genephase{\uset}{\smallcon}{\alpha}}}
{\intdu  e^{-\genephase{\uset}{\smallcon}{\alphatilde}}}
\end{equation}
In the formulation of Eq.~\ref{eq:deltaf} we are confronted with a
ratio of configurational integrals defined through the {\em same}
energy function (alias: `cost function' or `hamiltonian') acting on
{\em two explicitly different} regions of configuration space. In
contrast Eq.~\ref{eq:configintratio} features integrals defined through
{\em different} energy functions acting on {\em one common}
configuration space.

Configurational-integral ratios of the form of Eq.~\ref{eq:configintratio}
appear widely
in the free-energy literature;  but the underlying physical motivation
is not always the same. The spectrum of possible usages is covered by writing 
Eq.~\ref{eq:configintratio} in the more general form
\begin{equation}
\label{eq:configintratiogen}
{\cal R}_{A B}
= \frac
{\intdu  e^{-\genephasesimp{\uset}{A}}}
{\intdu  e^{-\genephasesimp{\uset}{B}}}
\equiv
\frac
{\genpfcalsimp_{A}}
{\genpfcalsimp_{B}}
\end{equation}
where $A$ and $B$ are two generalized macrostate labels, which 
identify two energy functions.

One meets this kind of ratio (perhaps most naturally) if one considers
the difference between the free energies of two macrostates of {\em
one} phase, corresponding to {\em different} choices of $\smallcon$ --
as in the influential work of Bennett \cite{bennett}. And one meets it
if one considers the difference between the free energies of a given
model and some approximation to that model, as in the perturbation
approach of Zwanzig \cite{zwanzig}. Rahman and Jacucci
\cite{rahmanjacucci} seem to have been amongst the first to consider
this structure of problem with the kind of motivation we have given it
here --that is, as a way of computing free energy differences between
two {\em phases} directly.

To set the remainder of the discussion in as wide a context as we can
we shall develop it in the general notation of Eq.~\ref{eq:configintratiogen},
reverting to the specific interpretation of the macrostate labels of 
primary interest here ($A \rightarrow \alpha, \smallcon$;
$B \rightarrow \alphatilde, \smallcon$) as appropriate.

In common with most who have addressed the problem posed by this kind
of configurational integral ratio, we shall exploit 
the statistics 
of an appropriate `order-parameter' defined on the common configuration 
space and of the general form
\begin{equation}
\label{eq:switchopgeneral}
\calop_{A B}
\equiv 
\switchopphase{M}{A}{\uset}
-\switchopphase{M}{B}{\uset}
\end{equation}
There is considerable license in the choice of
$\plainop_A$ and $\plainop_B$.
In the simplest cases it suffices to choose the two energy functions 
themselves.
We make that choice explicitly here so as to 
expose connections with  the work of others. Then the `order parameter' 
assumes the form
\begin{equation}
\label{eq:switchopenergy}
\calop_{A B}
=
\genephasesimp{\uset}{A}
-\genephasesimp{\uset}{B}
\end{equation}
This quantity has the credentials of an `order parameter' 
in the sense
that its behavior is qualitatively different in the two ensembles. To
understand this, suppose that we sample from the canonical distribution of 
the $B$-ensemble, prescribed by the partition function 
$\genpfcalsimp_{B}$
A `typical point' $\uset$ will then characterize a typical configuration
of $B$, in which --for example-- no particle penetrates the
core-region of the potential of another; that same point $\uset$
will, however, describe a configuration of $A$ which is {\em not}
guaranteed to be typical of that ensemble, and {\em will} in general
feature energy-costly regions of core penetration. The order parameter
$\calop_{A B}$ will thus generally be
{\em positive} in ensemble $B$; by the same token it will be {\em negative}
in ensemble $A$. 
(We shall see this explicitly in the examples that follow)

One may measure and utilize the statistics of $\calop$ in three 
strategically-different ways.
 
First, Eqs.~\ref{eq:configintratio} and \ref{eq:switchopenergy} lead 
immediately to the familiar Zwanzig formula \cite{zwanzig},
\begin{equation}
\label{eq:zwanzig}
{\cal R}_{AB}
= \expec{B}{e^{-\calop_{A B}}}
= \expec{B}{e^{-\left[\genephasesimp{\uset}{A} -\genephasesimp{\uset}{B}\right]}}
\end{equation}
In principle then {\em one single-ensemble-average}  
suffices to determine 
the desired ratio. However, this strategy (if unsupported by others)
requires \cite{rahmanjacucci} 
that the two ensembles overlap in the sense that, loosely \cite{moreprecisely},
the `dominant' configurations in the 
$A$ ensemble are a
subset of those in the $B$
ensemble.  This is a strong constraint; it will seldom if ever be 
satisfied \cite{butcanhelpwithtau}.

The second generic strategy \cite{bennett} utilizes {\em two single-ensemble-averages}, that is averages with respect to the {\em separate}
ensembles defined by 
$\genpfcalsimp_{A}$
and
$\genpfcalsimp_{B}$.
In particular one
may, in principle, measure the  canonical probability 
distributions
of the order
parameter in each ensemble separately,
and exploit the relationship between them \cite{rahmanjacucci}
\begin{equation}
\label{eq:pdfcondrel}
{\cal R}_{AB} P(\calop_{AB} \mid A)
= e^{-\calop_{AB}}
P(\calop_{AB} \mid B)
\end{equation}
again presupposing the choice prescribed in Eq.~\ref{eq:switchopenergy}.
One can re-express this result in an alternative form 
\begin{equation}
\label{eq:acceptanceformula}
{\cal R}_{AB} = 
\frac{
\expec{B}{\metrofunc{\calop_{AB} }}
}{
\expec{A}{\metrofunc{\calop_{BA} }}
}
\end{equation}
where $\metrofuncbare$ is the Metropolis function defined in
Eq.~\ref{eq:metrofuncdef}. Recalling the significance of that function
one can see each of the two terms on the RHS of Eq.~\ref{eq:acceptanceformula} as
the probability of acceptance of a Monte Carlo switch of the labels $A$ and $B$
(and thus of the controlling `hamiltonian') at a point in $\uset$-space, averaged over
that space. In the numerator this switch is from
$\genesimp_{B}$ to $\genesimp_{A}$ and the average is with
respect to the canonical distribution in ensemble $B$; in the
denominator the roles of $A$ and $B$ are reversed. This
is Bennett's acceptance ratio formula \cite{bennett}.

Eq.~\ref{eq:pdfcondrel} shows that for this strategy to work one needs
the two ensembles to `overlap' in the sense (somewhat less restrictive
than in the case of the Zwanzig formula, Eq.~\ref{eq:zwanzig}) that
the two single-ensemble \PDF s are measurable at some {\em common} value
of $\calop$, the most obvious candidate being the $\calop \simeq 0$
region intermediate between the values typical of the two ensembles.
Eq.~\ref{eq:acceptanceformula} shows that this requirement is
effectively equivalent to the condition that the probabilities of
acceptance of a hamiltonian switch can be measured, in {\em both}
directions.

In the form described, this strategy will virtually always fail; to
produce a generally workable strategy we need to introduce two further
ingredients, the first essential, the second desirable. First we need
to invoke ES techniques to extend the $\calop$-ranges sampled in the
single-ensemble simulations until they overlap; in principle this is
enough to allow us to determine the desired ratio by matching up the
two distributions in Eq.~\ref{eq:pdfcondrel}. But the information thus
gathered is more fully and efficiently utilized by taking a further
step. In the $\calop \simeq 0$ regions then accessed `switches'
(between the `hamiltonians' of the two ensembles) can be implemented not
just {\em virtually}, as envisaged in the acceptance ratio formula,
but {\em actually} within a configuration space enlarged to include
the macrostate label explicitly. One may then measure the full canonical
\PDF\ of $\calop \equiv \calop_{AB}$ and deduce the
desired ratio from (Eq~\ref{eq:configintratio})

\begin{equation}
\label{eq:switcheight}
{\cal R}_{AB}
=\frac{
\int_{\calop < 0} d{\cal M} \pcanon(\cal M | \smallcon) 
}{
\int_{\calop > 0} d{\cal M} \pcanon(\cal M |\smallcon) 
}
\end{equation}
where we have used the fact that the sign of the order parameter acts
as a signature of the ensemble.

This is the ESPS strategy. We have expressed it in a general way, 
as a switch between {\em any} nominated pair of `macrostates' or `ensembles'. 
In summarizing it, we revert to the particular context of primary interest
in which the two macrostates belong to {\em different} phases. 
The core idea is simple: to use ES methods
to seek out regions of the configuration space of one phase which are
such that a transition (switch, leap) to the other will be accepted
with reasonable probability. The leap avoids mixed-phase
configurations: the simulation explores both configuration spaces but
is always to be found in one or the other.

The full machinery of ESPS is customizable in a number of respects,
notably the choice of reference configurations, of order parameter and
of transformation matrix. These issues are best explored in the
context of specific examples.

\subsubsection{Examples \label{subsubsec:switchexamples}}

We consider three examples of ESPS, ordered conceptually rather than
chronologically.

\subsubsection*{Example A: hcp and fcc phases of Lennard Jones systems}

In the case of crystalline phases it is natural to choose the
reference configurations to represent the states of perfect crystalline order,
described by the appropriate sets of lattice vectors

\begin{equation}
\label{eq:choiceofqsetone}
\qsetref{\gamma} \longrightarrow \Rset{\gamma}  \hspace*{1cm} \gamma= \alpha, \alphatilde
\end{equation}

Note that the particle and lattice site indexing hidden in this
notation mandates a one-one mapping between the lattice sites of the
two systems; we are free to choose any of the $N!$ possibilities.
In the case of {\em hcp} and {\em fcc} lattices
\cite{useoflattice} it is natural to exploit the fact that the one
lattice can be made out of the other merely by translating
close-packed planes, in the fashion depicted in
Fig.~\ref{fig:lsfcchcp}.

The simplest choice for the $\taumat{}$ operations in
Eq.~\ref{eq:switchone} is to take them as the identity operation; the
operation $\smat$ (Eq.~\ref{eq:smat}) is then also the identity.

The choices of $\qsetref{}$ and $\taumat{}$ together define the geometry
of the switch operation - the way in which a configuration of one
structure is used to generate a configuration of the other: here the
switch exchanges one lattice for another, while conserving the
physical displacements with respect to lattice sites.

The final choice to be made is the form of the order parameter; in this case the
default (built out of the energy function) defined in
Eq.~\ref{eq:switchopenergy} proves the right choice. Thus, making the phase 
labels explicit we take
\begin{equation}
\label{eq:switchopLJ}
\calop_{\alpha \alphatilde}
=
\genephase{\uset}{\smallcon}{\alpha}
-\genephase{\uset}{\smallcon}{\alphatilde}
\end{equation}
 
Figures~\ref{fig:lsljpdf} and ~\ref{fig:lsljphasediag} show results
for the Lennard Jones (LJ) crystalline phases established with ESIT, on the basis of
these choices \cite{lssoftpot}. Commentary (of this and other results
here) is deferred to the `critique' below.

\subsubsection*{Example B: hcp and fcc phases of hard spheres}

We have already noted that the order parameter in ESPS need not be
constructed out of the true energy functions of the two phases. The
defining characteristic of an ESPS order parameter is that it measures
the difference between the values of some chosen function of the
common coordinate set $\uset$ evaluated in the two phases, such that
for some region (typically `sufficiently small values') of that
quantity, an inter-phase switch can be successfully initiated. An
ESPS `order parameter' merely provides a convenient thread that can be 
followed to the wormhole ends.

In the case of hard spheres the energy function does not provide a
usefully-graded measure of how far we are from a wormhole: the order
parameter defined in Eq.~\ref{eq:switchopenergy} will generally be
infinite because of hard-sphere overlap in the configurations created
by the switch. But it is easy to find an alternative:
all one has to do is build an order parameter
out of a count of the number $N^o$ of overlapping spheres. Instead of
Eq.~\ref{eq:switchopLJ} we then have
\begin{equation}
\label{eq:switchopHS}
\calop_{\alpha \alphatilde} 
\equiv
N^o_{\alpha}(\uset \, \smallcon)
-N^o_{\alphatilde}(\uset \, \smallcon)
\end{equation}
With the switch geometry chosen as for the LJ systems discussed above
the difference between the free energies of fcc and hcp hard sphere
systems can be determined precisely and transparently
\cite{lshsone,lshstwo}. Some of the results are included in
Table~\ref{table:hardspheredata}.

\subsubsection*{Example C: liquid and fcc phases of hard spheres}

The ESPS strategy can also be applied when one of the phases is a
liquid \cite{lsfreezing}. A configuration selected at random from
those explored in canonical sampling of the liquid phase will serve as
a reference state. Since the liquid and solid phases generally have
significantly different densities the simulation must be conducted at
constant pressure \cite{truealsoforcrystals}; the coordinate set
$\qset$ then contains the system volume and the switch must
accommodate an appropriate dilation (and can do so easily through the 
specification of the volumes implicit in the reference configurations).
While the overlap order parameter defined in
Eq.~\ref{eq:switchopHS} remains appropriate for simulations
conducted in the solid phase, in the liquid phase it is necessary to
engineer something a little more elaborate to account for the fact
that the particles are not spatially localized. 
Such considerations also lead to some relatively subtle
but significant finite-size effects. Figure~\ref{fig:lsfreezing} shows
some results locating the freezing pressure of hard spheres this way
\cite{lsfreezing}.

\subsubsection{Critique \label{subsubsec:switchcritique}}

The ESPS method draws on and synthesizes a number of ideas
in the extensive free energy literature,
including the importance of {\em representations} and 
{\em space transformations between them} \cite{rahmanjacucci,moody,voter}; 
the utility of {\em expanded ensembles} in turning virtual transitions into 
real ones \cite{lyub}; and the general power of {\em multicanonical
methods} to seek out macrostates with any desired property \cite{bn}.

Methodologically ESPS has much in common with ESIT: 
like ESIT it utilizes the paraphernalia of extended sampling to
visit both phases in a single simulation; but in contrast to ESIT it
contrives to do this {\em without} having to traverse the interfacial 
configurations which make ESIT hard, probably impossible, to implement
in problems 
involving solid phases.

ESPS thus shares a number of the advantages that ESIT has with respect
to  integration methods (Sections~\ref{subsubsec:integrationcritique}
\ref{subsubsec:traversecritique}).
It is pleasingly transparent: the evolution with temperature of
the relative stability of
{\em fcc} and {\em hcp} LJ crystals can be `read off' from
Fig.~\ref{fig:lsljpdf}; and the LJ freezing pressure `seen' in
Fig.~\ref{fig:lsfreezing}. Apart from finite-size effects uncertainties
are purely statistical. The fact that
both phases are realized within the same simulation means that
finite-size effects can be handled more systematically; this seems to be
a particular advantage of the ESPS approach to the liquid-solid phase boundary.

ESPS remains a computationally intensive strategy, though not
prohibitively so on the scale of its competitors:
one explicit comparison 
(in the case hard sphere crystals) indicates that ESPS and NIRM deliver
similar precision for similar compute resource \cite{pronkfrenkelone}.  

But the possibility of substantial improvements to ESPS remains. The
idea of improved (`targeted') mappings between the configuration
spaces has been discussed in general terms by Jarzynski
\cite{jarzynski}, albeit in the context of the Zwanzig formula
(Eq.~\ref{eq:zwanzig}) which will not generally work without ES-props.
In the ESPS framework that mapping enters in the matrix $\smat$,
reflecting the representations chosen for the two phases
(Eqs.~\ref{eq:smat},~\ref{eq:switchone}). One does not {\em have}
to preserve the physical displacements in the course of the switch.
By appropriate choice of the operations $\taumat{}$ it is possible
\cite{lsharmonic} to implement a switch which, instead, conserves a
set of Fourier coordinates, and thence the {\em harmonic}
contributions to the energy of the configurations of each phase; the
determinant in Eq.~\ref{eq:switchfour} then captures the harmonic
contribution to the free energy difference, leaving the computational
problem focused on the anharmonic contributions which (alone) are left
in ${\cal R}$. This strategy greatly enhances the 
overlap between the two branches of the order parameter
distribution; but the associated efficiency gains (resulting from
the reduced length of`path' through $\calop$-space) 
are offset by the greatly increased computational
cost of the mapping itself \cite{lsharmonic}.

\section{Doing it without paths: alternative strategies 
\label{SEC:DOINGWITHOUTPATHS}}

In this section we survey some of the strategic approaches to the
phase-coexistence problem which do not fit comfortably into the
path-based perspectives we have favored here. 

\subsection{The Gibbs Ensemble Monte Carlo method
\label{subsec:gibbsensemble}}

\subsubsection{Strategy \label{subsubsec:GEstrategy}}

Gibbs Ensemble Monte Carlo (GEMC) is an ingenious method introduced by
Panagiotopoulos\cite{PANAGIO87}, which allows one to simulate the
coexistence of liquid and vapor phases without having to deal with a
physical interface between them.

GEMC utilizes {\em two} simulation subsystems (`boxes'); though
physically separate the two boxes are thermodynamically coupled
through the MC algorithm which allows them to exchange both volume and
particles subject to the constraint that the total volume and number
of particles remain fixed. Implementing these updates (in a way that
respects detailed balance) ensures that the two systems will come to
equilibrium at a common temperature, pressure and chemical potential.
The temperature is fixed explicitly in the MC procedure; but the
procedure itself selects the chemical potential and pressure that
will secure equilibrium.

If the overall number density and temperature are chosen to lie within
the two-phase region, the system must phase separate; it does so
through configurations in which each box houses one pure
phase, since such arrangements avoid the free energy cost of an
interface.
The coexistence densities can then be simply measured through a sample
average over each box. By conducting simulations at a
series of temperatures, the phase diagram in the temperature-density
plane can be constructed. Details of the implementation
procedure can be found in reference~\cite{FRENKELSMIT}.

\subsubsection{Critique \label{subsubsec:GEcritique}}

The GEMC method is elegant in concept and simple in practice; it seems
fair to say that it revolutionized simulations of fluid phase
equilibria; it has been very widely used and comprehensively
reviewed \cite{PANAGIO2000}. We note three respects in which it is 
less than ideal.

First, in common with any simulation of open systems it runs into increasing
difficulties as one moves down the coexistence curve to the region of high 
densities where particle insertion (entailed by particle exchange) has a 
low acceptance probability.

It also runs into difficulties of a different kind at the other end of
the coexistence curve, as one approaches the critical point. GEMC
effectively supposes that criticality may be identified by the
coalescence of the two peaks in the separate branches of the density
distribution captured by the two simulation boxes. The limiting
critical behavior of the full density distribution in a system of
finite size (Fig.~\ref{fig:nbwljcp}, to be discussed below) 
shows that this is not so;
the critical point cannot be reliably located this way. These difficulties
are reflected in the strong finite-size-dependence of the
shape of the coexistence curve evident in GEMC studies
\cite{VALLEAU98}. To  make sense of the GEMC behavior near the 
critical point, therefore, one needs
to invoke finite-size scaling strategies \cite{monbind,BRUCE97}; but
these are substantially less easy to implement than they are in the
framework of the $\mu VT$ ensemble, to be discussed in 
section~\ref{subsec:finitesize}.

Finally we note that the efficiency of GEMC is reduced through the
high computational cost of volume moves (each one of which requires a
recalculation of {\em all} inter-particle interactions). Comparisons
show \cite{PANAGIO2000} that the strategy discussed in
Sec~\ref{subsubsec:traversestrategy} (multicanonical methods and
histogram re-weighting, within a grand canonical ensemble) gives a
better return in terms of precision per computational unit cost. But
GEMC is undoubtedly easier to implement.

\subsection{The NPT \& test particle method \label{subsec:NPT-TP}}

\subsubsection{Strategy \label{subsubsec:NPT-TPstrategy}}

The NPT-TP method \cite{NPT_TEST} locates phase coexistence at a prescribed
temperature by finding that value of the pressure for which the
chemical potentials of the two phases are equal.
The chemical potential $\mu$ is identified with the difference between the 
Helmholtz free energies of systems containing $N$ and $N-1$ particles. Then 
the Zwanzig formula (\cite{zwanzig}, Eq.~\ref{eq:zwanzig}) shows that
\begin{equation}
\label{eq:widom}
\mu = \ln \expec{N-1}{e^{-\left[\genesimp_{N} -\genesimp_{N-1} \right]}}
= \mu^{ig} + \ln \expec{N-1}{e^{-\Delta U_{N,N-1}}}
\end{equation}
where $\Delta U_{N,N-1}$ is the additional configuration energy 
associated with the insertion of a `test'
particle into a system of $N-1$ particles.  Equation
~\ref{eq:widom} is due to Widom 
\cite{WIDOM} and forms the basis of the
{\em test particle insertion method}. 
In contrast to other realizations of the Zwanzig formula it works 
(or can do so) reasonably well, because the argument of the exponential  
is of $\order{1}$ rather than $\order{N}$, as long as the particle 
interactions are short-ranged.

The values of the chemical potentials in each phase, together with
their pressure-derivatives (available through the measured number
densities) can be exploited to home in on the coexistence
pressure. The method has been successfully applied to calculate the
phase diagrams of a number of simple fluids and fluid mixtures
\cite{LOTFI,VRABEC}.

\subsubsection{Critique \label{subsubsec:NPT-TPcritique}}

The NPT-TP method is obviously designed to deal with the coexistence
of fluid phases: particle-insertion into ordered structures is
generally to be avoided. In this restricted context it is
straightforward to implement, needing no more than the conventional
apparatus of single-phase NPT simulation. However in common with other
strategies that involve particle insertion (such as Gibbs Duhem
integration, Sec.~\ref{subsec:tracking}) it runs into difficulties at
high densities; and it cannot readily handle the behavior in the
critical region (Sec.~\ref{subsubsec:critical}).

\subsection{Beyond equilibrium sampling: fast growth methods \label{subsec:FGmethods}}

\subsubsection{Strategy \label{subsubsec:FGstrategy}}

The techniques discussed in this section might reasonably have been
included in our collection of path-based strategies
(Section~\ref{SEC:PATHBASEDTECHNIQUES}). However, although the idea of a
`path' features here too, it does so {\em without} the usual
implications of equilibrium sampling.

At the heart of the techniques in question (we shall refer to them 
collectively as {\em Fast Growth}, FG)
is a simple and beautiful result 
established by Jarzynski \cite{jznefd} which we write in the 
form \cite{factorsofT}
\begin{equation}
\label{eq:jznefd}
{\cal R}_{AB}
\equiv  e^{-\Delta {\cal F}_{AB}}
= \overline {e^{-W_{AB}^{\tau_S}}}
\end{equation}
Here $W_{AB}^{\tau_s}$ is the work done
in switching the effective energy function (through some time-dependent 
control parameter $\lambda(t)$)
from $\genesimp _B$ to
$\genesimp _A$, in time $\tau_s$, while the system observes the 
dynamical or stochastic updating rules appropriate to the 
energy function appropriate at any instant.
The bar denotes an average over the ensemble of such procedures
generated by choosing the initiating microstate randomly from
macrostate $B$. 

Equation~\ref{eq:jznefd} incorporates two more familiar claims as
special cases. In the limit of {\em long} switching times the system has
time to equilibrate at every stage of the switching procedure; then
Eq.~\ref{eq:jznefd} reduces to the result of numerical integration
along the path prescribed by the switching operation (cf Eq.~\ref{eq:lambdaint})
\begin{equation}
\label{eq:jznefdint}
\Delta {\cal F}_{AB}
= \int_{B}^{A} d \lambda \expec{\lambda}
{\frac{\partial \genesimp}
{\partial \lambda}
}
\end{equation}
At the other extreme, if the switching time $\tau_s$ is short, 
the work done is just the energy cost of an instantaneous and 
complete hamiltonian switch, 
and one recovers the Zwanzig formula (cf Eq.~\ref{eq:zwanzig})
\begin{equation}
\label{eq:jznefdzwanzig}
\Delta {\cal F}_{AB}
= -\ln \expec{B}{e^{-\left[\genesimp_{A} -\genesimp_{B}\right] }}
\end{equation}
In fact Eq.~\ref{eq:jznefd} holds {\em irrespective} of the switching
time $\tau_s$: the {\em equilibrium} free-energy difference is
determined by the spectrum of a quantity, $W_{AB}$, associated with a
{\em non-equilibrium} process. The `exponential average' 
of the work done in taking the system between the designated macrostates
(at {\em any} chosen rate) thus provides an alternative estimator of the difference between the associated free energies.

The fact that Eq.~\ref{eq:jznefd} (in general) folds in non-equilibrium
processes is made explicit through a third result which may be deduced from it.
The convexity of the exponential \cite{convexityofexp} implies that 
\begin{equation}
\label{eq:exponentialav}
\overline{e^{-W_{AB}}} \ge e^{-\overline{W_{AB}}}
\end{equation}
so that, from Eq.~\ref{eq:jznefd},
\begin{equation}
\label{eq:jznefdinequality}
\Delta {\cal F}_{AB} \le \overline{W_{AB}}
\end{equation}
This is the Helmholtz inequality, a variant of the Second Law
and thus an acknowledgment of the consequences of irreversible processes.

\subsubsection{Critique \label{subsubsec:FGcritique}}

Our discussion of the Zwanzig formula shows that the FG representation
is unlikely to be practically helpful if one chooses small $\tau_s$
--at least for the systems having large enough $N$ to be of interest to us
here. But given a choice of devoting a specified computational resource to
{\em one long} switch (and appealing to the integration procedure,
Eq.~\ref{eq:jznefdint}) or {\em several short} switches (and appealing
to the exponential averaging procedure Eq.~\ref{eq:jznefd}) the latter
seems to be the preferred strategy
\cite{jznefd}: the two approaches are comparable in precision; but
FG generates
an estimate of its own uncertainties; and it is trivially implementable
in parallel computing architectures. 
This seems a potentially fruitful avenue for 
further exploration. 

One can avoid the issues associated with exponential-averaging if one
uses the FG formula in the form of the inequality
Eq~\ref{eq:jznefdinequality}, in tandem with the corresponding
inequality emerging from a reverse switch operation (from A to B).
The two results together give upper and lower bounds on the difference
between the two free energies; the bounds can be tightened 
by a variational procedure with respect to the parameters of the 
chosen switch \cite{twoboundsvarop}. Figure~\ref{fig:millrein} shows the 
results of such a procedure \cite{millrein} applied to the Bain 
switch operation \cite{bain} 
which maps an {\em fcc} lattice onto a 
{\em bcc} lattice by a continuous \cite{baincomment} deformation.

\section{Determining the phase boundary: extrapolation, tracking and the 
thermodynamic limit \label{SEC:RESTOFTHEJOB}}}

The path-based methods exemplified in the preceding section provide us
with ways of estimating the difference between the free energies of
the two phases, $\Delta \genf{\alphaalphatilde}{\lambdaset, N}$
(Eq.~\ref{eq:deltaf}) at {\em some} point $\smallcon \equiv \lambdaset$ in
the space of the controlling fields \cite{carewithsets},
for a system whose size $N$ (which we shall occasionally make
explicit in this section) is computationally manageable.

The practical task of interest here requires that we identify the set
of points $\smallcon_x \equiv \lambdaset_x$ giving phase coexistence in 
the thermodynamic
limit, and defined by the solutions to the equation
\begin{equation}
\label{eq:coexcond}
\lim_{N\rightarrow \infty} 
\frac{1}{N}\Delta \genf{\alphaalphatilde}{\lambdaset_x, N}
=0
\end{equation}
We divide this programme into three parts. The first issue is how to
use the data accumulated at our chosen state point to infer the
location of {\em some} point $\smallcon_x $ on the coexistence curve, for 
some finite $N$.
The
second is to map out the phase boundary emanating from that point. And
the third is to deal with the corrections associated with
the limited (`finite') size of the simulation system. 

We shall restrict the discussion to a two-dimensional field space 
spanned by fields $\lambdaset\equiv \lambda_1, \lambda_2 $ with conjugate
macrovariables $\plainopset \equiv \plainop_1, \plainop_2$. 

\subsection{Extrapolation to the phase boundary \label{subsec:extrapolation}}

The simplest way of using the measurements at $\lambdaset$ to estimate the
location of a point on the phase boundary is to perform a linear 
extrapolation in one of the fields using the result
\begin{equation}
\label{eq:conjugates}
\frac{\partial \Delta\genf{\alphaalphatilde}{\lambdaset}}{\partial \lambda} = 
\expec{\alpha}{\plainop} -\expec{\alphatilde}{\plainop}
\end{equation}

Note that the quantities on the RHS of this equation represent
separate single-phase expectation values, defined with respect to
single-phase canonical distributions of the form given in
Eq.~\ref{eq:partialdistdef}; they are thus problem-free.

The implied estimate of the coexistence-value of the field
$\lambda_1$ (say) is then
\begin{equation}
\lambda_{1x} =\lambda_1 - 
\frac{\Delta \genf{\alphaalphatilde}{\lambdaset}}{
\expec{\alpha}{\plainop_1} -\expec{\alphatilde}{\plainop_1}}
\end{equation}
Such extrapolations provide a simple, but possibly crude, way
of correcting an initially poor estimate of coexistence. 

One may be able to do better by appealing to histogram re-weighting
techniques (Appendix~\ref{APP:HISTOGRAMREWEIGHTING}). {\em If} the initial
measurement of a free energy difference is based on some form of
extended sampling which establishes the full canonical distribution
of some order parameter; and {\em if} that order parameter is the
conjugate of one of the fields spanning the phase diagram of interest;
{\em then} HR allows one to scan through a range of values of that
field to find the coexistence value (identified by the resulting
equality of the areas of the two peaks in the canonical order
parameter distribution, implied by Eq.~\ref{eq:deltaf}). 
Such a process is easily automated.

\subsection{Tracking the phase boundary \label{subsec:tracking}}

In principle knowledge of a {\em single point} on the coexistence curve
permits the {\em entire curve} to be traced without further calculation of 
free energies. The key result needed follows
simply from Eq.~\ref{eq:conjugates}, applied to 
each field $\lambda_1$ and $\lambda_2$ in turn. 
The {\em slope} of the coexistence curve
$\Delta \genf{\alphaalphatilde}{\lambdaset_x}=0$, 
follows as
\begin{equation}
\label{eq:genclauclap}
\left[\frac{d\lambda_1}{d\lambda_2}\right]_x =- \frac
{\expec{\alpha}{\plainop_2} -\expec{\alphatilde}{\plainop_2}}
{\expec{\alpha}{\plainop_1} -\expec{\alphatilde}{\plainop_1}}
\end{equation}
This is the generalized Clausius-Clapeyron equation 
\cite{reducestocc}. It expresses the slope entirely in terms of
single-phase averages; the slope can
be employed (in a predictor-corrector
scheme) to estimate a nearby coexistence point. Fresh simulations
performed at this new point yield the phase boundary gradient there,
allowing further extrapolation to be made, and so on. In this manner one
can in principle track the whole coexistence curve. This strategy is widely
known as Gibbs-Duhem integration (GDI) \cite{KOFKE}.

GDI has been used effectively in a number of studies, most notably in
the context of freezing of hard and soft spheres
\cite{GDIexamples}. Its distinctive feature is simultaneously its
strength and its weakness: once boot-strapped by knowledge of one point
on the coexistence curve it subsequently 
requires only {\em single-phase} averages.
This is clearly a virtue since the elaborate machinery needed for 
two-phase sampling is not to be unleashed lightly.  But {\em without} any
`reconnection' of the two configuration-spaces at subsequent simulation state
points, the GDI approach offers no feedback on integration
errors. Since there will generally exist a {\em band} of practically
stable states on each side of the phase boundary, it is possible for
the integration to wander significantly from the true boundary
with no indication that anything is wrong.

A more robust (though computationally more intensive) alternative to
GDI is provided by a synthesis of extended (multicanonical) sampling
and histogram re-weighting techniques. The method is boot-strapped by an
ES measurement of the full canonical distribution of a suitable
order parameter, at some point on the coexistence curve (identified by
the equal areas criterion specified in Eq.~\ref{eq:deltaf}). HR
techniques then allow one to map a region of the phase boundary close
to this point. The range over which such extrapolations are reliable
is limited (Appendix~\ref{APP:HISTOGRAMREWEIGHTING}) and it is not
possible to extrapolate arbitrarily far along the phase boundary:
further multicanonical simulations will be needed at points that lie
at the extremes of the range of reliable extrapolation.  But there is
no need to determine a new set of weights (a new extended sampling
distribution) from {\it scratch} for these new simulations. HR allows
one to generate a {\em rough estimate} of the equilibrium order
parameter (at these points) by extrapolation from the original measured
distribution. The `rough estimate' is enough to furnish a usable set
of weights 
for the new multicanonical simulations.
Repeating the combined procedure (multicanonical simulation
followed by histogram extrapolation) one can track along the coexistence
curve. The data from the separate histograms can subsequently be
combined self consistently (through
multihistogram extrapolation, as discussed in Appendix~\ref{APP:HISTOGRAMREWEIGHTING} ) 
to yield the
whole phase boundary. If one wishes to
implement this procedure for a phase boundary that terminates in a
critical point it is advisable to start the tracking procedure nearby.
At such a point 
the ergodic block presented to inter-phase traverses
is relatively small (the canonical order parameter distribution is
relatively weakly doubly-peaked); and so the multicanonical distribution 
(weights) required to initiate the whole process can be determined without 
extensive (perhaps without any) iterative procedures \cite{getthis}. 

\subsection{Finite-size effects \label{subsec:finitesize}}

Computer simulation is invariably conducted on a model system whose
size is small on the thermodynamic scale one typically has in mind
when one refers to `phase diagrams'. Any simulation-based study of
phase behavior thus necessarily requires careful consideration of 
`finite-size effects'. The nature of these effects is significantly 
different according to whether one is concerned with behavior close to or 
remote from a critical point. The distinction reflects the relative sizes
of the linear dimension $L$ of the system --the edge of the simulation 
cube-- and the correlation length $\xi$ --the distance over which the 
local configurational variables are correlated. By `non-critical' we 
mean a system for which $L \gg \xi$; by critical we mean one for which
$L \ll \xi$. We shall discuss these two
regions in turn, and avoid the lacuna in between.

\subsubsection{Non-critical systems}

In the case of non-critical systems the issue of finite-size effects
is traditionally expressed in terms of the finite-size corrections to
the free energy densities of each of the two phases:
\begin{equation}
\label{eq:fscs}
f_\alpha(\smallcon, \infty) - f_\alpha(\smallcon, N) 
\hspace*{0.5cm}
\mbox{where}
\hspace*{0.5cm}
f_\alpha(\smallcon, N) =\frac{1}{N} \genf{\alpha}{\smallcon, N}
\end{equation}
In recent years substantial efforts have been made to develop a
theoretical framework for understanding the nature of such corrections
\cite{privmanbook}. In the case of {\em lattice models} (ie models of 
strictly localized particles) in the {\em $NVT$-ensemble} with {\em
periodic boundary conditions} (PBCs) it has been established {\em a
priori} \cite{borgtheory} and corroborated in explicit simulation
\cite{borgexperiment} that the corrections are {\em exponentially small}
in the system size \cite{equalareas}.

However these results do not immediately carry
over to the problems of interest here where (while PBCs are the norm)
the ensembles are frequently open or constant pressure, and the
systems do not fit in to the lattice model framework. Even in the 
apparently simple case of crystalline solids in $NVT$,
the free translation of the center of mass
introduces $N$-dependent phase space-factors 
in the configurational integral which manifest themselves
as additional finite-size corrections to the free energy; these
may not yet be fully  understood \cite{polson,lsfreezing}. 
If one adopts the traditional stance then, one is typically faced with 
having to make extrapolations of the free energy densities
in {\em each} of the two phases, {\em without} a secure understanding of the
underlying form ($\frac{1}{N}$? $\frac{\ln N}{N}$? \ldots) of the 
corrections involved.

The problems are reduced if one shifts the focus of attention from the
single-phase free energies to the quantities of real interest: the
difference between the two free energies, and the field-values that
identify where it vanishes. In both cases within an ES strategy that
treats both phases together, there is only {\em one} extrapolation to
do, which is clearly a step forward.  If the two phases are of the
same generic type (eg two crystalline solids) one can expect
cancellations of corrections of the form $\frac{\ln N}{N}$ --which
should be identical in both phases-- leaving presumably at most
$\frac{1}{N}$ corrections. If the phases are not of the same generic
type (a solid and a liquid) the logarithmic corrections will probably
not cancel; reliable extrapolation will be possible only if they can be
identified and allowed for explicitly, leaving only a pure power law.

Overall it seems that there is considerable room for progress here.

\subsubsection{Near-critical systems \label{subsubsec:critical}}

In the case of simulation studies of near-critical systems, the issues
associated with `finite-size effects' have an altogether different
flavor. First, they are no longer properly regarded as essentially
small effects to be `corrected for'; the critical region is
characterized by a strong and distinctive dependence of
system-properties on system size; the right strategy is to address
that dependence head on, and exploit it. Second, there is an
extensive framework to appeal to here: the phenomenology of {\em
finite-size scaling} \cite{privmanbook} 
and the underpinning theoretical structure of the
{\em renormalization group} \cite{RG} together show what to look for and what to
expect. Third, the objectives are rather different, going well beyond the 
issues of phase-diagram mapping that we are preoccupied with here.

Our discussion will be substantially briefer than it might be; we will focus
on the issue most relevant here (but not not the most interesting)
-- the location of the  critical point in a fluid. 

As in the coexistence curve problem, the key is the distribution
of the order parameter, in this case the density. On the coexistence
curve, remote from the critical point (ie in the region $\xi\ll L$) we
have seen (Fig.~\ref{fig:nbwlj}) that this distribution comprises two
peaks of equal area, each roughly centered on the corresponding single
phase average; the two peaks are narrow and near-Gaussian in form (the
more so the larger the system size) as one would expect from the
Central Limit Theorem; the probability of inter-phase tunneling (an
inverse measure of the ergodic barrier) is vanishingly small
\cite{howbigisit}. As one moves up the coexistence curve simulations 
show more or less what one would guess: the peaks broaden; they become
less convincingly Gaussian; and the tunneling probability increases:
this is the natural evolution en route to the form which must be
appropriate in the one-phase region beyond criticality -- a single peak
narrowing with increasing $L$, asymptotically Gaussian.  
Against this immediately
intelligible backdrop, the behavior at criticality
(Fig.~\ref{fig:nbwljcp}) is comparatively subtle. 
Again (for large enough $L$, but still $L\ll \xi$) a limiting form is reached;
however that limiting form comprises neither one Gaussian nor two, but 
something in between. The distribution narrows with increasing $L$
(while preserving the shape of the `limiting form'); however here
the width varies not as $1/L^{d/2} =1/\sqrt{N}$
(familiar from Central Limit behavior) but as $1/L^{\beta/\nu}$
where $\beta$ and $\nu$ are the critical exponents of the order parameter 
and the correlation length respectively. There is good reason to believe that
the shape of the distribution shares the distinctive quality of the 
critical exponents -- all are `universal signatures of behavior common to 
a wide range of physically disparate systems. The idea (of long-standing 
\cite{originsofuniversality}) 
that fluids and Ising magnets belong to the same universality class is 
corroborated (beyond the exponent level) by the correspondence between 
their critical-point order parameter distributions \cite{adbnbw}.
Once that correspondence 
has been established to ones satisfaction one is  at liberty to 
{\em exploit} it to refine the assignment of fluid critical-point parameters.
The form of the density distribution depends sensitively on
the choice for the controlling fields (chemical potential $\mu$ and 
temperature $T$); {\em demanding} 
correspondence with the form appropriate for the Ising universality class 
(studied extensively and known with considerable precision)  
sets narrow bounds on the location of the fluid critical point \cite{nbwlj}.

Finally we should not overlook one -perhaps unexpected- feature of
the critical distribution: it shows that clear signatures of the two
(incipient) phases persist along the coexistence curve and {\em right
through to the critical point} in a finite system. Failure to appreciate this
will lead (as it has in the past) to substantial overestimates of
critical temperatures.

\section{Dealing with imperfection \label{SEC:IMPERFECTION}}

Real substances often deviate from the idealized models employed in
simulation studies. For instance many complex fluids, whether natural
or synthetic in origin, comprise mixtures of {\em similar} rather than
{\em identical} constituents. Similarly, crystalline phases usually
exhibit a finite concentration of defects which disturb the otherwise
perfect crystalline order. The presence of imperfections can 
significantly alter phase behavior with respect to the idealized case.
If one is to realize the goal of obtaining quantitatively accurate
simulation data for real substances, the effects of imperfections must
be incorporated. In this section we consider the state-of-the-art in
dealing with two kinds of imperfections, polydispersity and point
defects in crystals.

\subsection{Polydispersity} 

Statistical mechanics was originally formulated to describe the
properties of systems of {\em identical} particles such as atoms or
small molecules. However, many materials of industrial and commercial
importance do not fit neatly into this framework. For example, the
particles in a colloidal suspension are never strictly identical to
one another, but have a range of radii (and possibly surface charges,
shapes etc). This dependence of the particle properties on one or more
continuous parameters is known as polydispersity. One can regard a
polydisperse fluid as a mixture of an infinite number of distinct
particle species. If we label each species according to the value of
its polydisperse attribute, $\sigma$, the state of a polydisperse
system entails specification of a density {\em distribution} $\rho(\sigma)$,
rather than a finite number of density variables. It is usual to
identify two distinct types of polydispersity: {\em variable} and {\em
fixed}. Variable polydispersity pertains to systems such as ionic
micelles or oil-water emulsions, where the degree of polydispersity (as
measured by the form of $\rho(\sigma)$) can change under the influence
of external factors. A more common situation is fixed
polydispersity, appropriate for the description of systems such as
colloidal dispersions, liquid crystals and polymers. Here the form of
$\rho(\sigma)$ is determined by the synthesis of the fluid. 

Computationally, polydispersity is best handled within a grand
canonical (GCE) or semi-grand canonical ensemble in which the density
distribution $\rho(\sigma)$ is controlled by a conjugate chemical
potential distribution $\mu(\sigma)$. Use of such an ensemble is
attractive because it allows $\rho(\sigma)$ to fluctuate as a whole,
thereby sampling many different realizations of the disorder and hence
reducing finite-size effects. Within such a framework, the case of
variable polydispersity is considerably easier to tackle than fixed
polydispersity: the phase behavior is simply obtained as a function of
the width of the prescribed $\mu(\sigma)$ distribution. Perhaps for
this reason, most simulation studies of phase behavior
in polydisperse systems have focused on the variable case
\cite{STAPLETON,KOFKE,KRISTOF,THEODOROU}. 

Handling fixed polydispersity is computationally much more challenging:
one wishes to retain the efficiency of the GCE, but to do so, a way
must be found to adapt the imposed form of $\mu(\sigma)$ such as to
realize the prescribed form of $\rho(\sigma)$. This task is
complicated by the fact that $\rho(\sigma)$ is a {\em functional} of
$\mu(\sigma)$. Recently, however, a new approach has been developed which
handles this difficulty. The key idea is that the
required form of $\mu(\sigma)$ is obtainable iteratively by
functionally minimizing a cost function quantifying the deviation of
the measured form of $\rho(\sigma)$ from the prescribed `target' form. For
efficiency reasons, this minimization is embedded within a histogram
re-weighting scheme (Appendix~\ref{APP:HISTOGRAMREWEIGHTING})
obviating the need for a new simulation at
each iteration. The new method is efficient, as evidenced by tests on
polydisperse hard spheres \cite{WILDING02} where it permitted the first
direct simulation measurements of the equation of state of a
polydisperse fluid.

\subsection{Crystalline defects}

Defects in crystals are known to have a potentially major influence on
phase behavior. For instance, dislocation unbinding is believed to be
central to the 2D melting transition, while in 3D there is evidence to
suggest that defects can act as nucleation centers for the liquid phase
\cite{GOMEZ}.  In superconductors, defects can pin vortices and
influence `vortex melting' \cite{RUDNEV}

Almost all computational studies of defect free-energies (and their
influence on phase transitions) have been concerned with point defects.
Polson {\em et al} \cite{polson}, used the results of early
calculations \cite{BENNET71} of the vacancy free energy of a hard
sphere crystal, to estimate the equilibrium vacancy concentration at
melting. Comparison with the measured free energies of the perfect hard
sphere crystal (obtained from the Einstein Crystal NIRM method discussed in
Sec~\ref{subsec:integration}) 
allowed them to estimate the effect of vacancies on
the melting pressure, predicting a significant shift.  A
separate calculation for interstitials found their equilibrium
concentration at melting to be 3 orders of magnitude smaller than that
of vacancies. In follow-up work, Pronk and Frenkel \cite{pronkfrenkeltwo} 
used a technique similar to the Widom particle insertion method 
(Section~\ref{subsubsec:NPT-TPstrategy})
to calculate the
vacancy free energy of a hard sphere crystal. For interstitial defects 
they employed an extended sampling technique in which a tagged `ghost
particle' is grown reversibly in an interstitial site.

To our knowledge there have been no reported measurements of equilibrium
defect concentrations in soft spheres models. Similarly,  relatively
few measurements have been reported of defect free energies in models
for real systems. Those that exist rely on  integration methods to
connect the defective solid to the perfect solid. In ab-initio studies
the computational cost of this procedure can be high, although results
have recently started to appear, most notably for vacancies and
interstitial defects in Silicon. For a review see reference \cite{ALFE}.

\section{Outlook \label{SEC:outlook}}

Those who read this paper may share with its authors the feelings expressed in
reference \cite{ecclesiastes}: the dynamics in this particular
problem space seems to have been rather more diffusive than ballistic. 
It is therefore wise to have some idea of where the ultimate destination is, and the strategies that are most likely to take us there.

The long term goal is a computational framework that will be grounded
in electronic structure as distinct from phenomenological particle
potentials; that will predict global phase behavior {\em a priori},
rather than simply decide between two nominated candidate phases; and
that will handle quantum behavior, in contrast to the essentially
classical framework on which we have focused here. That goal is
distant but not altogether out of sight. Integrating {\em ab-initio}
electronic structure calculations with the statistical mechanics of
phase behavior has already received some attention
\cite{ALFE,gjareview}. The WL algorithm
(\cite{WANGLAN}, Appendix~\ref{APP:WEIGHTS})
offers a glimpse of the kind of 
self-monitoring configuration-space search algorithm that one needs
to make automated {\em a priori} predictions of phase behavior 
possible. And folding in quantum mechanics requires only a dimensionality 
upgrade \cite{qft}.

There are of course many other challenges, a little less grand: the
two space and time scales arising in asymmetrical binary mixtures
\cite{binmix}; the fast attrition (exponential in the chain length)
for the insertion of polymers in NPT-TP or grand-canonical 
methods \cite{longmols};
the long range interactions in coulombic fluids \cite{coulombfluids};
the extended equilibration times for dense liquids near the
structural glass transition \cite{KOB00}; and the extreme long-time-dynamics 
of the escape from metastable states in nanoscale ferromagnets
\cite{KOLESIK02}

As regards the strategies that seem most likely to take us forward, we
make three general observations:

Making the most of the information available in simulation studies
requires an understanding of finite-size effects; it also requires
awareness of the utility of quantities that one would not naturally
consider were one restricted to pen and paper.

We will surely need new algorithms;
they come from physical insight into the configurational core of the
problem at hand.

One needs to match formulations to the available technology: parallel
computing architectures give some algorithms a head-start.

\appendix

\renewcommand{\theequation}{\Alph{section}\arabic{equation}}

\section{Building extended sampling distributions
\label{APP:WEIGHTS}}

In contrast to canonical sampling distributions whose form can be
{\em written down} (Eq.~\ref{eq:gendistdef}) the Extended Sampling
(ES) distributions discussed in
section~\ref{subsubsec:extendedsampling} have to be {\em
built}. There is a large literature devoted to the building
techniques, extending back at least as far as reference 
\cite{torrvalleau}. We
restrict our attention to relatively recent developments (those that
seem to be reflected in current practices); and we shall focus on
those aspects which are most relevant to ES distributions facilitating
two-phase sampling.

In the broad-brush classification scheme offered in
Sec.~\ref{subsubsec:extendedsampling} the domain of an ES distribution
may be prescribed by a range of values of one or more {\em fields}, or
one or more {\em macrovariables}. We shall focus on the latter
representation which seems simpler to manage. The generic task then is
to construct a (`multicanonical') sampling distribution which will
visit all (equal-sized) intervals within a chosen range of the
nominated macrovariable(s) with roughly equal probability: the multicanonical
distribution of the  macrovariable(s) is essentially flat over the 
chosen range.

In formulating a strategy for building such a distribution, most
authors have chosen to consider the particular case in which the
macrovariable space is one-dimensional, and is spanned by the
configurational {\em energy}, $E$ \cite{useofE}. The choice is
motivated by the fact that a distribution that is multicanonical in
$E$ samples configurations typical of a range of {\em temperatures},
providing access to the simple (reference-state) behavior that often
sets in at high or low temperatures. For the purposes of two-phase
sampling we typically need to track a path defined on some
macrovariable other than the energy (ideally, in {\em addition} to it: 
we will come back to this). The hallmarks of a `good' choice are that in some
region of the chosen variable (inevitably one with intrinsically low
equilibrium probability) the system may pass (has a workably-large
chance of passing) from one phase to the other.  In discussing the key
issues, then, we shall have in mind this kind of quantity; we shall
continue to refer to it as an order parameter, and denote it by
$\plainop$.

One can easily identify the generic structure of the sampling distribution 
we require. It must be of the form
\begin{equation}
\label{eq:ESAPPsampdist}
\psamp(\qset) \eqdot \frac {\pcanon (\qset\mid \smallcon)}
{\hat{\pcanon}(\plainop (\qset))}
\end{equation}
Here $\hat \pcanon (\plainop)$ is an {\em estimate} of the true canonical 
$\plainop$-distribution. Appealing to the sampling identity~\ref{eq:identity}
the resulting $\plainop$-distribution is
\begin{equation}
\label{eq:ESAPPopdist}
\psamp( \plainop ) =
\expec{S}{\delta[\plainop -\plainop(\qset)]}
\eqdot 
\expec{\equm}{\frac{\psamp}{\pcanon}\delta[\plainop -\plainop(\qset)]}
\eqdot \frac
{\pcanon(\plainop \mid\smallcon)}
{\hat{\pcanon}(\plainop) }
\end{equation}
and is multicanonical (flat) to the extent that our estimate of the canonical
$\plainop$-distribution is a good one. 

We can also immediately write down the prescription for generating
the ensemble of configurations defined by the chosen
sampling distribution. We need a simple MC procedure with acceptance 
probability (Eq.~\ref{eq:metropacc}, with the presumption of 
Eq.~\ref{eq:ptrialsymmetry})
\begin{equation}
\label{eq:metropaccsimp}
\pacc (\qset \rightarrow \qsetprime) = 
\metrobare{
\frac
{\psamp(\qsetprime)}
{\psamp(\qset)} 
}
\end{equation}

In turning this skeleton framework into a working technique one must make
choices in regard to three key issues:
\begin{enumerate}
\item How to {\em parameterize} the estimator $\hat{\pcanon}$.
\item What {\em statistics} of the ensemble to use to guide the
update of $\hat{\pcanon}$.
\item What {\em algorithm} to use in updating $\hat{\pcanon}$.
\end{enumerate}

The second and third issues are the ones of real substance; the issue
of parameterization is important only because the proliferation of
different choices that have been made here may give the impression
that there are more techniques available than is actually the
case. That proliferation is due, in some measure, to the preoccupation
with building ES distributions for the {\em energy}, $E$. 
There are as many `natural
parameterizations' here as there are ways in which $E$ appears in
canonical sampling.  Thus Berg and Neuhaus
\cite{bn} employ an $E$-dependent{\em effective temperature};
Lee \cite{LEE93} utilizes a {\em microcanonical entropy} function;
Wang and Landau \cite{WANGLAN} focus on a {\em density of
states} function. Given our concern with macrovariables 
other than $E$ the most appropriate parameterisation of the
sampling distribution here is through a {\em multicanonical weight function} 
$\mucw (\plainop)$, in practice represented by a discrete set of
multicanonical weights $\{\mucw \}$. 
Thus we write \cite{signconvention}
\begin{equation}
\label{eq:ESAPPweightparamone}
\hat{\pcanon} (\plainop \mid \smallcon) \eqdot e^{-\mucw (\plainop)}
\end{equation}
implying (through Eq.~\ref{eq:ESAPPsampdist}) a sampling distribution
\begin{equation}
\label{eq:ESAPPweightparamsamp}
\psamp(\qset) \eqdot \sum_{j=1}^{\Omega} e^{- \genesimp (\qset) -\mucw _j}
\deltapick{j}{\plainop}
\eqdot e^{- \beta E (\qset) +\mucw \left[\plainop(\qset)\right]}
\end{equation}
which is of the general form of Eq.~\ref{eq:extendedsamplingdensity}, with
$w_j\equiv e^{\mucw _j}$.

There are broadly two strategic responses to the second of the issues
raised above: to drive the estimator in the right direction one may
appeal to the {\em statistics of visits to macrostates} or to the {\em
statistics of transitions between macrostates}. We divide our
discussion accordingly.

\subsection{Statistics of visits to macrostates} 

The extent to which any chosen sampling distribution (weight function)
meets our requirements is reflected most directly in the
$\plainop$-distribution it implies. One can estimate that distribution
from a histogram $\hist (\plainop)$ of the macrostates visited in the
course of a set of MC observations. One can then use this information
to refine the sampling distribution to be used in the next set of MC
observations. The simplest update algorithm is of the
form \cite{SMITH95}
\begin{equation}
\label{eq:ESAPPweightupdate}
\hat{\eta} (\plainop) 
\longrightarrow \
\hat{\eta}(\plainop )-\ln[\hist(\plainop)+1] +k
\end{equation}
The form of the logarithmic term serves to ensure that macrostates
registering no counts  (of which there will usually be
many) have their weights incremented by
the same finite amount (the positive constant $k$ \cite{choiceofk}); 
macrostates which {\em have} been visited are (comparatively) down-weighted.

Each successive iteration comprises a fresh simulation, performed
using the weight function yielded by its predecessor; since the
weights attached to unvisited macrostates is enhanced (by $k$) at
every iteration which fails to reach them, the algorithm plumbs a
depth of probablility that grows exponentially with the iteration number.  The
iterations proceed until the sampling distribution is
roughly flat over the entire range of interest.

There are many tricks of the trade here. One must recognize the
interplay between signal and noise in the histograms: the algorithm
will converge only as long as the signal is clear.  To promote faster
convergence one can perform a linear extrapolation from the
sampled into the unsampled region. One may bootstrap the process by
choosing an initial setting for the weight function on the basis of
results established on a smaller (computationally-less-demanding)
system. To avoid spending excessive time sampling regions in
which the weight function has already been reliably determined, one can
adopt a multistage approach. Here one determines the weight
function separately within slightly overlapping windows of the
macrovariable. The individual parts of the weight function are then
synthesized using multi-histogram re-weighting (Appendix~\ref{APP:HISTOGRAMREWEIGHTING}) to obtain the full
weight function. For further details the reader is referred to
\cite{bergrev}.

The strategy we have discussed is generally attributed to Berg and Neuhaus (BN)
\cite{bn}. Wang and Landau (WL) have offered an alternative formulation
\cite{WANGLAN}. To expose what is different, and what is not, it is
helpful to consider first the case in which the macrovariable is the 
energy, $E$. Appealing to what one knows {\it a priori} about the
canonical energy distribution, the obvious parameterization
is
\begin{equation}
\label{eq:ESAPPweightparamtwo}
\hat{\pcanon} (E) \eqdot \hat{\dos }(E) e^{-\beta E}
\end{equation}
where $\hat{\dos }(E)$ is an estimator of the {\em density of states} function 
$\dos (E)$.
Matching this parameterization to the multicanonical weight function
$\hat{\eta}(E)$ implied by choosing $M=E$ in
Eq.~\ref{eq:ESAPPweightparamone} one obtains the correspondence
\begin{equation}
\label{eq:BNupdate}
\hat{\eta}(E) = \beta E - \ln \hat{\dos } (E)
\end{equation}
There is thus no major difference here.
The differences between the two strategies
reside rather in the procedure by which the parameters of the sampling 
distribution are {\em updated}, and the point at which that procedure
is {\em terminated}.

Like BN, WL monitors visits to macrostates. But, while BN
updates the weights of {\em all} macrostates
after {\em many} MC steps, WL updates its `density
of states' for {\em the current} macrostate
after {\em every} step. The update prescription is
\begin{equation} 
\label{eq:WLupdate}
\hat{\dos }(E) \longrightarrow f \hat{\dos }(E)
\end{equation}
where $f$ is a constant, greater than unity. As in BN a visit to a
given macrostate tends to reduce the probability of further
visits. But in WL this change takes place {\em immediately} so the sampling
distribution evolves on the basic timescale of the simulation.  As the
simulation proceeds, the evolution in the sampling distribution irons
out large differences in the sampling probability across $E$-space,
which is monitored through a histogram $\hist(E)$. When that
histogram satisfies a nominated `flatness criterion' the entire process
is repeated (starting from the current $\hat{\dos }(E)$, but zeroing
$H(E)$) with a smaller value of the weight-modification factor,
$f$.  

Like BN, then, the WL strategy entails a two-time scale iterative
process. But in BN the aim is only to generate a set of weights that
can be utilized in a further, final multicanonical sampling process; the
iterative procedure is terminated when the weights are sufficiently
good to allow this. In contrast, in WL the iterative procedure is
pursued further --to a point \cite{howfardoesitgo}  where
$\hat{\dos }(E)$ may be regarded as a definitive approximation to $\dos (E)$,
which can be used to compute any (single phase)
thermal property at any temperature through
the partition function
\begin{equation} 
\label{eq:dostopf}
\genpfsimp (\beta) = \int dE \, \dos (E) e^{-\beta E}
\end{equation}

In the context of energy sampling, then, BN and WL achieve 
essentially the same ends, by algorithmically-different routes. Both
entail choices (in regard to their update schedule) which have to
rest on experience rather than any deep understanding. WL seems
closer to the self-monitoring ideal, and may scale more favorably
with system size.

The WL procedure {\em can} be applied to any chosen macrovariable,
$\plainop$. But while a good estimate $\hat{\dos }(E)$ is sufficient to
allow multicanonical sampling in $E$ (and a definitive one 
is enough to determine $\genpfsimp (\beta)$, Eq.~\ref{eq:dostopf})
the $\plainop$-density of states does not itself deliver the desired
analogues: we need, rather, the {\em joint} density of states
$\dos (E,\plainop)$ which determines the restricted,
single-phase partition functions through
\begin{equation} 
\label{eq:jointdostopf}
\genpfsimp_\alpha (\beta) = \int dE \int dM \,\dos (E, \plainop) 
e^{-\beta E} \deltapick{\alpha}{\plainop}  
\end{equation}
The WL strategy {\em does} readily generalize to a 2D macrovariable space. 
The substantially greater investment of computational resources is 
offset by the fact that the relative weights of the two phases can be 
determined at {\em any} temperature. Reference~\cite{twodESdistsexample}
provides one of (as yet few) illustrations of this strategy, which seems 
simple, powerful and general.

\subsection{Statistics of transitions between macrostates} 

The principal general feature of the algorithms based on visited macrostates is 
that the domain of the macrovariable they explore expands relatively slowly 
into the regions of interest. The algorithms we now discuss offer 
significant improvement in this respect. Although (inevitably, it seems)
they exist in a variety of guises, they have a common core which
is easily established. We take
the general detailed balance condition 
Eq.~\ref{eq:detailedbalance} and sum over configurations $\qset$ and 
$\qsetprime$ that contribute (respectively) to the  
macrostates $\plainop_i$ and $\plainop_j$ of some chosen 
macrovariable $\plainop$.
We obtain immediately
\begin{equation}
\label{eq:macrodetailedbalance}
\psamp (\plainop_i)
\overline {\psamp (\plainop_i \rightarrow \plainop _j)}
=
\psamp (\plainop_j)
\overline {\psamp (\plainop_j \rightarrow \plainop _i)}
\end{equation}
The terms with over-bars are macrostate transition probabilities (TP). 
Specifically $\overline {\psamp (\plainop_i \rightarrow \plainop _j)}$ is the 
probability (per unit time, say) of a transition from some nominated 
configuration in $\plainop_i$ to {\em any} configuration in $\plainop_j$, 
ensemble-averaged over the configuration in $\plainop_i$. 
Adopting a more concise (and suggestive) matrix notation:
\begin{equation}
\label{eq:macrodetailedbalancematrix}
\pvec{\samp}{\plainop}{i} \pmat{\samp}{\plainop}{ij}
=
\pvec{\samp}{\plainop}{j} \pmat{\samp}{\plainop}{ji}
\end{equation}
This is a not-so-detailed balance condition; 
it holds for any sampling distribution and any macrovariable \cite{SMITH95}. 
The components of the
eigenvector of the TP matrix (of eigenvalue unity) thus identify
the macrostate probabilities. This is more useful than it might seem.
One can build up an approximation of the transition matrix by  monitoring 
the transitions which follow when a simulation is launched from an 
{\em arbitrary} point in configuration space. The `arbitrary' point can be 
judiciously sited in the heart of the interesting region; 
the subsequent simulations then carry the macrovariable right though the 
chosen region, allowing one to accumulate information about it from the outset.
With a sampling distribution 
parameterized  as in Eq.~\ref{eq:ESAPPweightparamsamp}, the 
update scheme is simply
\begin{equation}
\label{eq:ESAPPweightupdateTP}
\hat{\eta} (\plainop) \longrightarrow \hat{\eta}(\plainop )-\ln
[\hat{\psamp} (\plainop)] +k
\end{equation}
where $\hat{\psamp} (\plainop)$ is the estimate of the sampling distribution 
that is {\em deduced} from the measured TP matrix \cite{caveat}. Reference
~\cite{grsadbspt} describes the application of this technique to
a structural phase transition.

One particular case of Equation~\ref{eq:macrodetailedbalance} has
attracted considerable  attention. If one sets $\plainop =E$, and
considers the infinite temperature limit, the probabilities of the
macrostates $E_i$ and $E_j$ can be replaced by the associated values
of the density of states function $\dos (E_i)$ and $\dos (E_j)$ The
resulting equation has been christened {\em the broad-histogram
relation} \cite{broadhistogram}; it forms the core of extensive
studies of transition probability methods referred to
variously as `flat histogram' \cite{flathistogram} and `transition
matrix' \cite{transitionmatrix}. Applications of these formulations
seem to have been restricted to the situation where the 
energy is the macrovariable, and the energy spectrum is discrete.

Methods utilizing macrostate transitions do have one notable advantage
with respect to those that rely on histograms of macrostate-visits. In
transition-methods the results of separate simulation runs 
(possibly initiated from different points in macrovariable space)
can be straightforwardly combined: one simply aggregates the 
contributions to the transition-count matrix \cite{SMITH95}. Synthesizing 
the information in  separate histograms (see the succeeding Appendix)
is less straightforward. The easy synthesis of data sets makes the TP method 
ideally suited for implementation in parallel architectures. 
Whether these advantages are sufficient to offset the 
the fact that TP methods are undoubtedly more complicated to implement
is, perhaps, a matter of individual taste.

\section{Histogram re-weighting \label{APP:HISTOGRAMREWEIGHTING}}

Without loss of generality the effective configurational energy may 
always be written in the form
\begin{equation}
\label{eq:hrone}
\genesimp(\qset, \lambdaset) = 
- \sum_{\mu} \lambda^{(\mu)} \plainop ^{(\mu)}(\qset)
\end{equation}
where $\lambdaset$ and $\plainopset$ are sets \cite{carewithsets}
comprising one or more mutually 
conjugate fields and macrovariables \cite{hiddenparameters}.
We consider two
ensembles which differ only in the values of one or more of the  fields 
$\lambdaset$.
The canonical 
sampling distributions of the two ensembles are then related by
\begin{equation}
\label{eq:hrtwo}
\pcanon(\qset | \lambdasetprime)
e^{-\sum_{\mu} \lambda^{\prime}\, ^{(\mu)} \plainop ^{(\mu)}(\qset)}
\eqdot
\pcanon(\qset | \lambdaset)
e^{-\sum_{\mu} \lambda ^{(\mu)} \plainop ^{(\mu)}(\qset)}
\end{equation}
where, again, 
$\eqdot$ signifies equality to within a configuration-independent constant.
Performing the configurational sum, for {\em fixed} values of the 
macrovariables $\plainopset$ then yields the relationship
\begin{equation}
\label{eq:hrthree}
\pcanon(\plainopset | \lambdasetprime)
e^{-\sum_{\mu} \lambda^{\prime}\, ^{(\mu)} \plainop ^{(\mu)}}
\eqdot
\pcanon(\plainopset | \lambdaset)
e^{-\sum_{\mu} \lambda ^{(\mu)} \plainop ^{(\mu)}}
\end{equation}
so that
\begin{equation}
\label{eq:hrfour}
\pcanon(\plainopset | \lambdasetprime)
\eqdot
\pcanon(\plainopset | \lambdaset)
e^{-\sum_{\mu} [\lambda ^{(\mu)} -\lambda^{\prime}\,^{(\mu)}] \plainop ^{(\mu)}}
\end{equation}
{\em In principle} then 
the $\plainopset$-distribution for {\em any} values of
the fields can be inferred from the 
$\plainopset$-distribution
for {\em one} particular set of values
\cite{theconstant}.
This is the basis of the Histogram Re-weighting (HR) technique
\cite{FERRENBERG}, also known as Histogram Extrapolation 
\cite{whyhistogram}.
It can be seen as 
a special case of the sampling identity given in Eq.~\ref{eq:identity}
\cite{hrfromidentity}. Like that identity its formal promise is not  
matched by what it can deliver in practice. The measurements in the
$\lambdaset$-ensemble (from which the extrapolation is to be made)
determine only an {\em estimate} of the $\plainopset$ distribution for
that ensemble. The estimate will be relatively good for the most
probable $\plainopset$ values (around the `peak' of that distribution)
which are `well-sampled', and relatively poor for the less probable
values (in the `wings'), which are sampled less well. This trade-off
(desirable if one wants only properties of the $\lambdaset$ ensemble)
limits the range of field-space over which extrapolation will be
reliable. The distribution associated with a set $\lambdasetprime$
remote from $\lambda$ may peak in a region of $\plainopset$-space far
from the peak of the measured distribution, lying instead in its
poorly sampled wings. In such circumstances the estimate provided by
the extrapolation prescribed by Eq.~\ref{eq:hrfour} will be
unreliable (and will typically reveal itself as such, through its 
ragged appearance).
 
In the scheme we have discussed extrapolations are made on the basis
of a histogram determined at a {\em single} point in field-space. The
multi-histogram method \cite{FERRENBERG} extends this framework.
It entails a sequence of separate simulations spanning a range of the
field (or fields) whose conjugate macrovariable(s) are of interest.
The intervals are chosen so that the tails of the histograms of the
macrovariable accumulated at neighboring state points overlap.  It is
clear from the discussion above that in principle {\em every}
histogram will provide some information about {\em every} region; and
that the most reliable information about any {\em given} region will come
from the histogram which samples that region most effectively. These
ideas can be expressed in an explicit prescription for synthesizing
all the histograms to give an estimate of the canonical distribution
of the macrovariable across the whole range of 
fields \cite{FERRENBERG,similarto}.

\epsfclipon

\newpage

\newpage

\newpage

\newpage

\newpage
\begin{figure}
\setlength{\epsfxsize}{14.0cm}
\centerline{\mbox{\epsffile{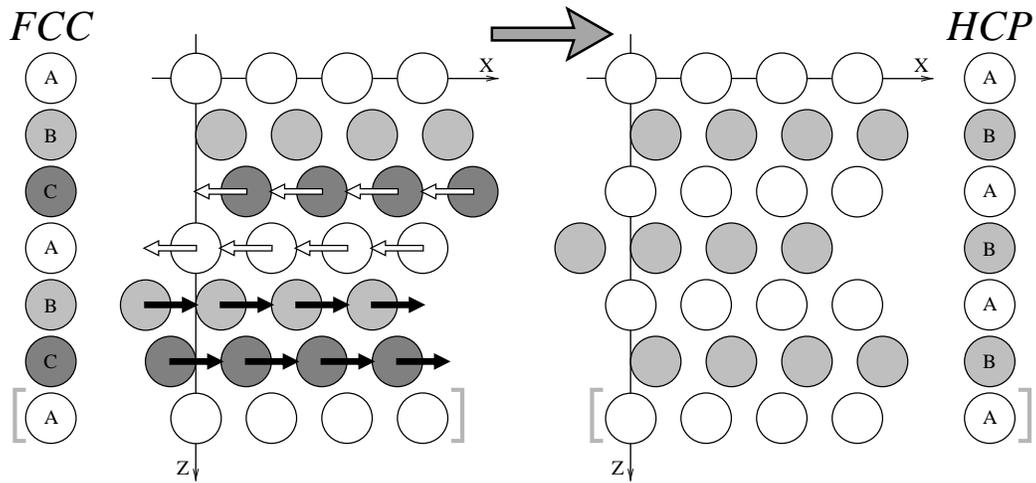}}} 

\vspace*{1cm}

\caption{
\label{fig:lsfcchcp}
The simple transformation for switching between {\em fcc} and {\em hcp} 
lattices.
The diagram shows 6 close-packed ($x-y$) layers. 
(The additional bracketed layer
at the bottom is the periodic image of the layer at the top.)
The circles show the boundaries of particles located 
at the sites of the two  close-packed structures. 
In the lattice switch operation
the top pair of planes are left unaltered, 
while the other pairs
of planes are relocated by translations, specified by 
the black and white arrows. The switch operation is discrete: the relocations
occur `through the wormholes'.
\protect\figcite{4}{lshstwo}
}

\end{figure}

\newpage
\begin{figure}
\setlength{\epsfxsize}{14.0cm}
\centerline{\mbox{\epsffile{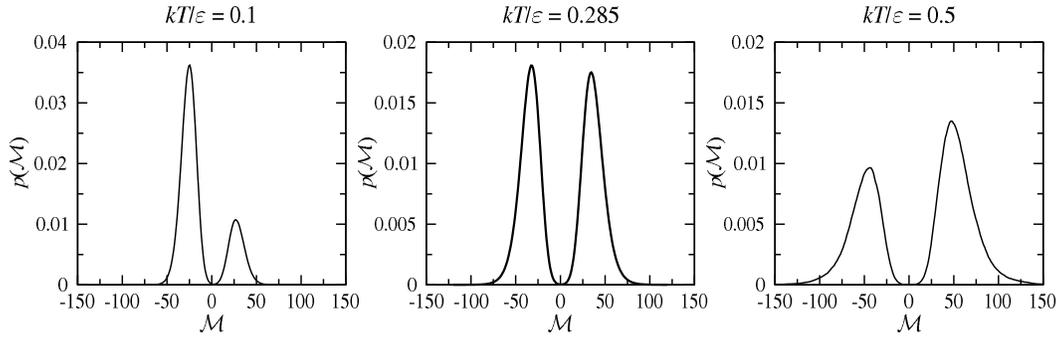}}} 

\vspace*{1cm}

\caption{
ESPS (`Lattice Switch') studies of the relative stability of {\it
fcc} and {\it hcp} phases of the LJ solid at zero pressure, as discussed in
section~\ref{subsec:switch}. In
this case the order parameter $\plainop$ (Eq.~\ref{eq:switchopLJ})
measures the difference between the energy of a configuration of one
phase and the corresponding configuration of the other phase generated
by the switch operation.  The areas under the two peaks reflect the
relative configurational weights of the two phases. The evolution with
increasing temperature (from {\it hcp}-favored to {\it fcc}-favored
behavior) picks out the {\it hcp}-{\it fcc} phase boundary shown in
Fig.~\ref{fig:lsljphasediag}.
\protect\figcite{7}{lssoftpot}
}
\label{fig:lsljpdf} 
\end{figure}

\newpage
\begin{figure}
\setlength{\epsfxsize}{14.0cm}
\centerline{\mbox{\epsffile{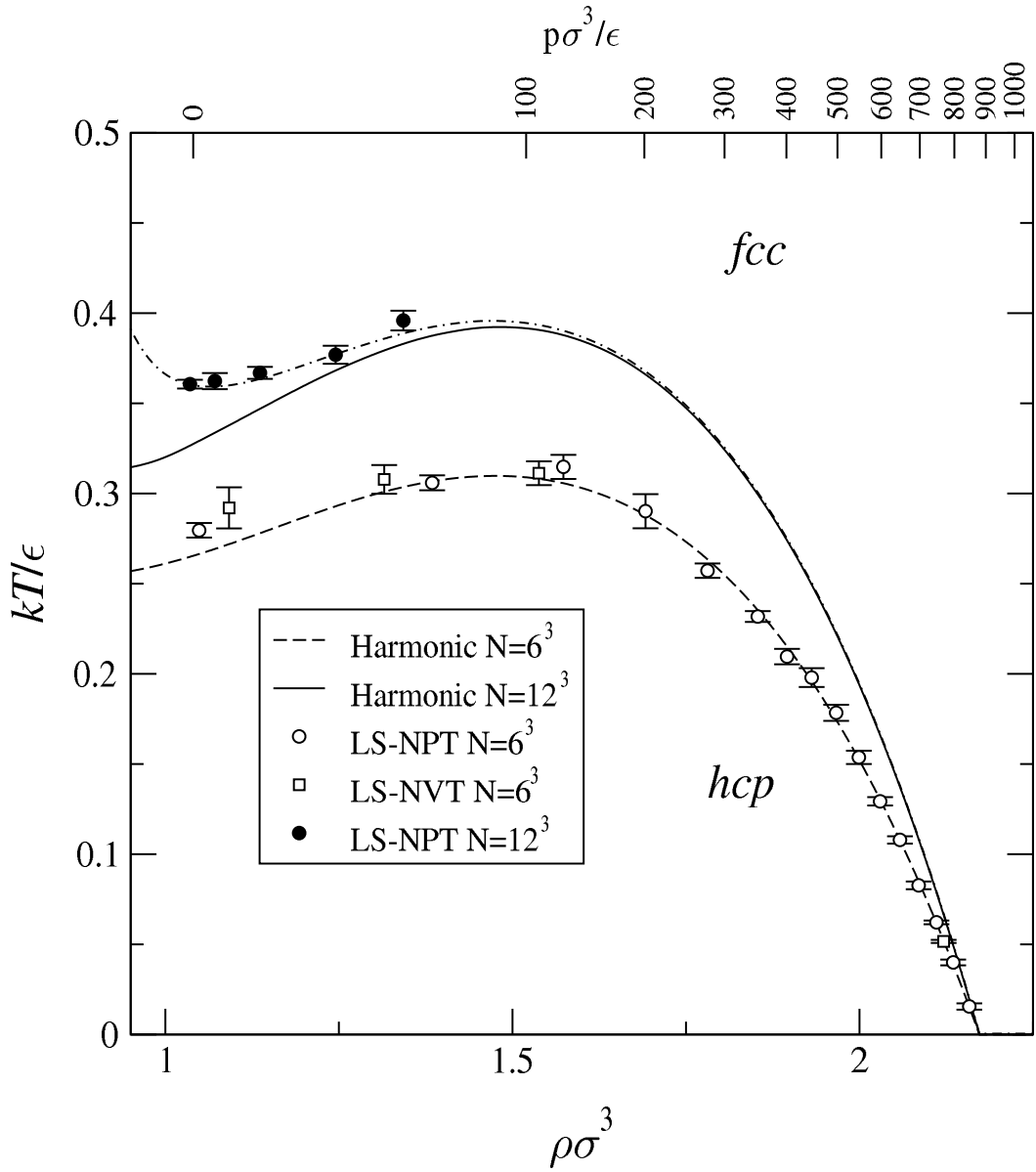}}} 

\vspace*{1cm}

\caption{
A variety of approximations to the classical Lennard Jones
phase diagram. The data points show the results of ESPS 
studies (discussed in section~\ref{subsec:switch}),  
denoted here by `LS'.
The dashed and solid lines are the 
results of harmonic calculations (for the two system sizes).
The dash-dotted line is a phenomenological parameterisation of the 
anharmonic effects 
The scale at the top of the figure shows the pressures at selected
points on the (LS $N=12^3$, $NPT$) coexistence curve.
Tie-line structure is unresolvable on the scale of the figure .
\protect\figcite{11}{lssoftpot} }
\label{fig:lsljphasediag}
\end{figure}

\newpage
\begin{figure}
\setlength{\epsfxsize}{14.0cm}
\centerline{\mbox{\epsffile{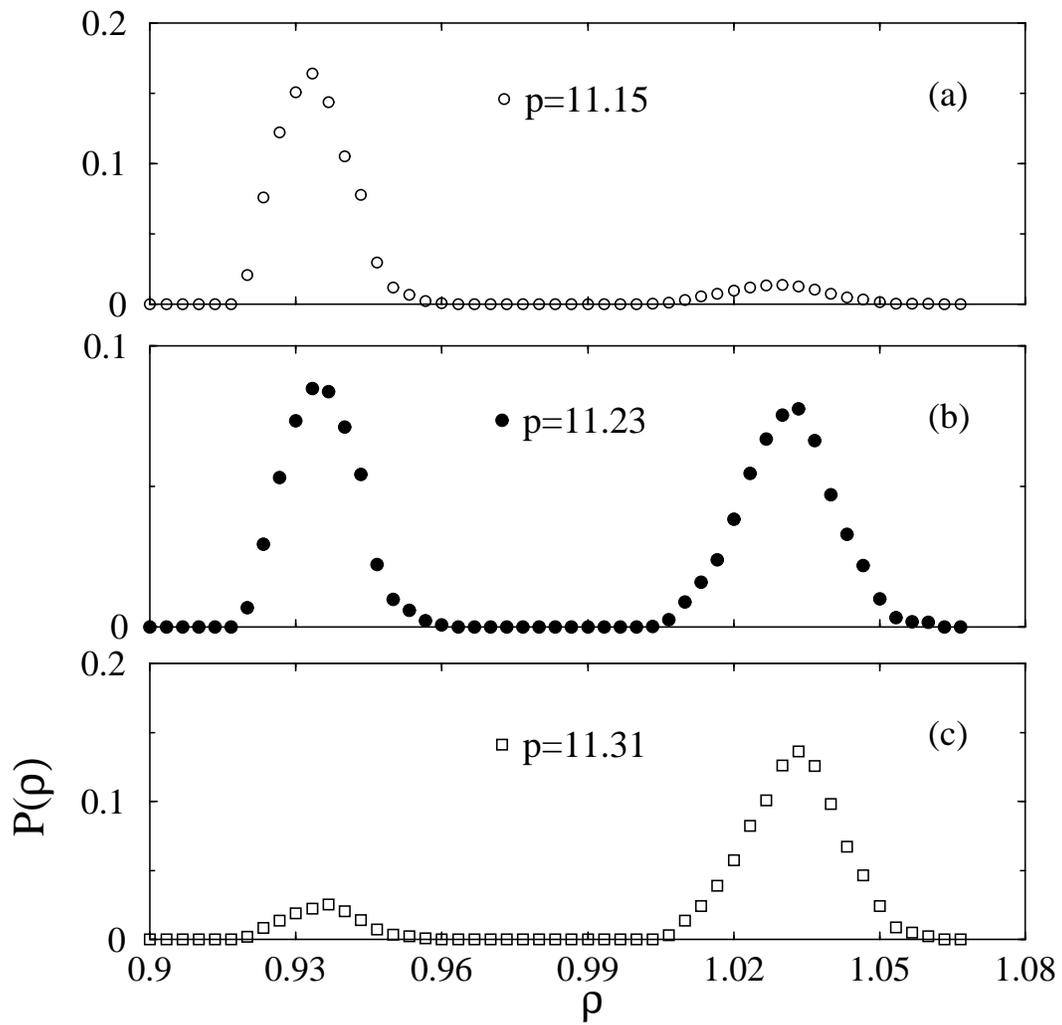}}} 

\vspace*{1cm}

\caption{
The distribution of the density of the system of $N=256$ LJ particles in
crystalline and liquid phases, as determined by ESPS methods. 
The three pressures are (a) just below, (b) at and  (c) just above 
coexistence for this $N$. 
\protect\figcite{1}{lsfreezing}
}
\label{fig:lsfreezing}
\end{figure}

\newpage
\begin{figure}
\setlength{\epsfxsize}{14.0cm}
\centerline{\mbox{\epsffile{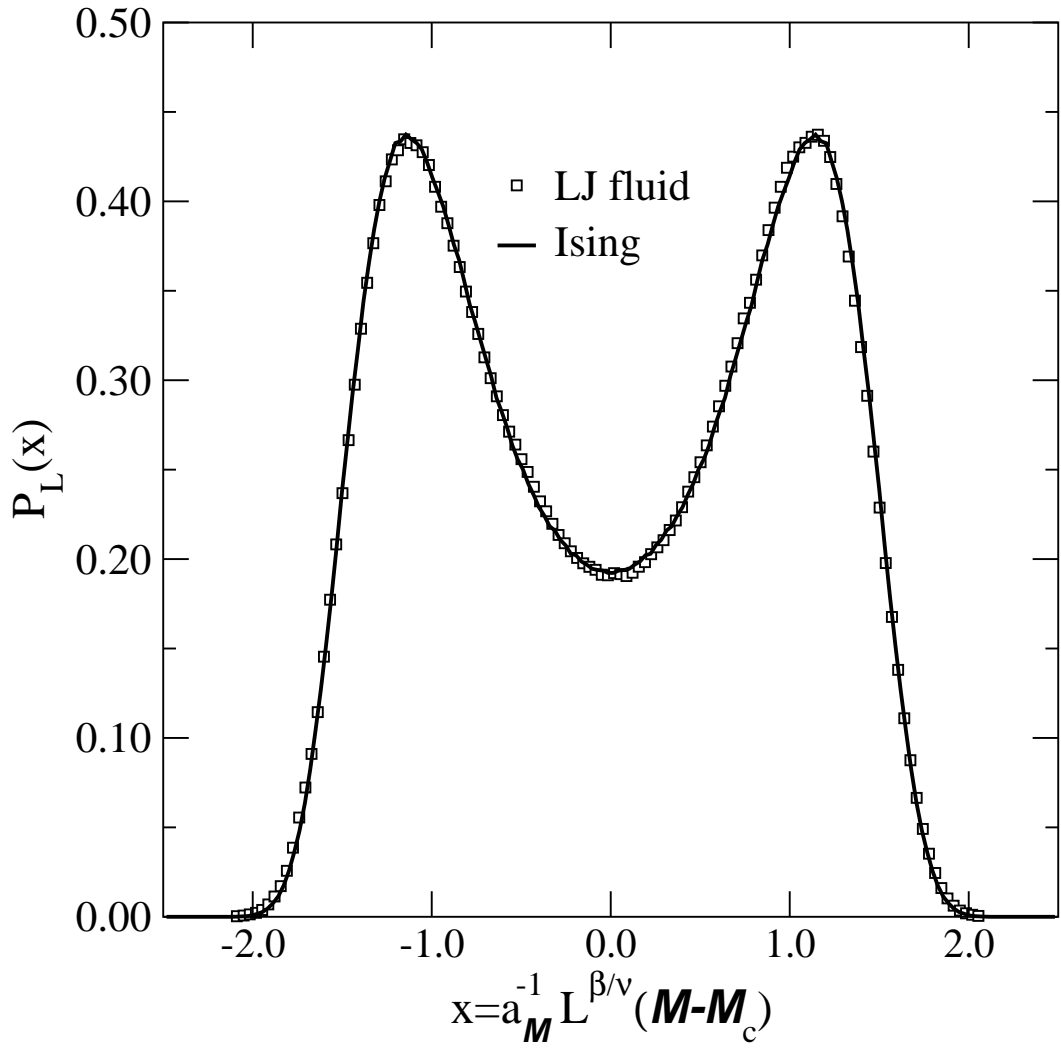}}} 

\vspace*{1cm}

\caption{
The distribution of the density of a LJ fluid at its critical point 
showing the collapse (given a suitable choice of scale) onto
a form characteristic of the Ising universality class.
\protect\figcite{3a}{nbwlj}}

\label{fig:nbwljcp}
\end{figure}

\newpage
\begin{figure}
\setlength{\epsfxsize}{14.0cm}
\centerline{\mbox{\epsffile{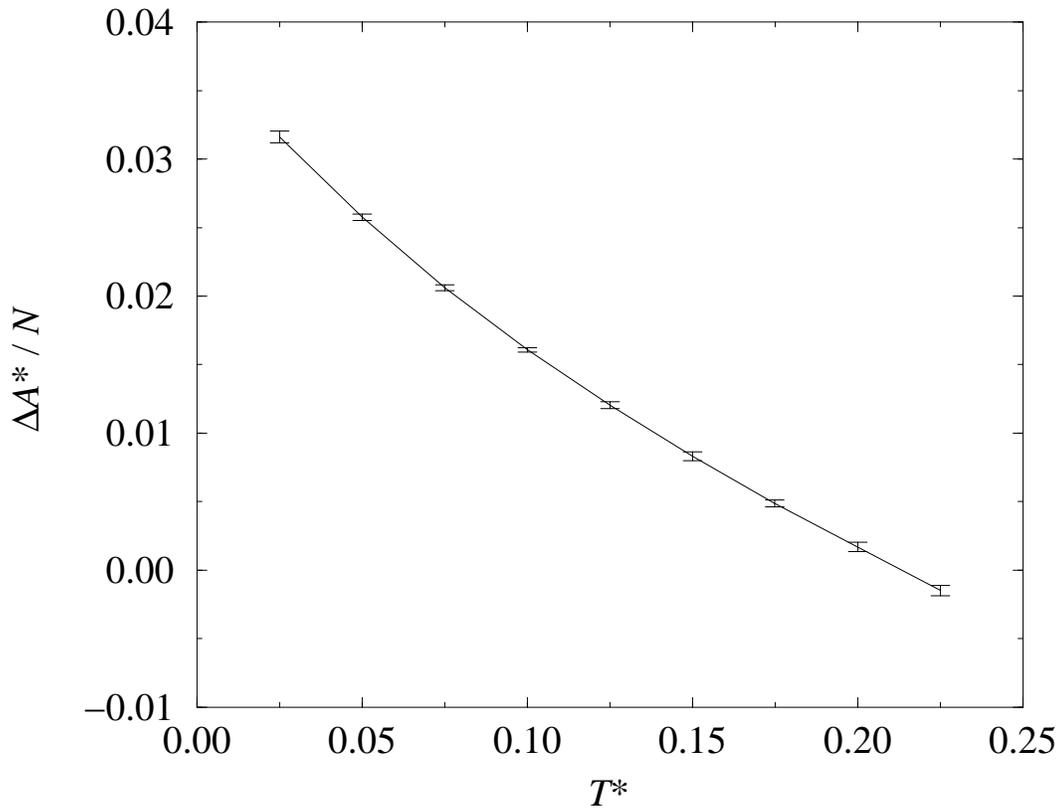}}} 

\vspace*{1cm}

\caption{
The difference between the free energy densities of {\it fcc} and {\it bcc}
phases of particles interacting through a Yukawa potential, as a function of
temperature, determined through the FG methods discussed in section
~\ref{subsec:FGmethods}. The error bars reflect the difference between the
upper and lower bounds provided by FG-switches between the phases 
(along the Bain-path \protect\cite{bain}) in the two directions. 
The favored phase changes around $T^{\star}= 0.21$.
\protect\figcite{11}{millrein}
}

\label{fig:millrein}
\end{figure}

\newpage

\begin{table}[htb!]
    \begin{center}
    \begin{tabular}[h]{rrrrrr}
      $\rho / \rho_{cp}$ & N & \multicolumn{2}{c}{$\Delta s$  $(10^{-5}\times k_B )$} & Method & Ref.\\
      0.736  & 12000 &  230 & (100) &   NIRM &\cite{woodcock}\\
      0.736  & 12096 &   87 &  (20) &    NIRM &\cite{bolfrenkelmh}\\
      0.7778 &   216 &  132 &   (4) &    ESPS &\cite{pronkfrenkelone}\\
      0.7778 &  1728 &  112 &   (4) &    ESPS &\cite{pronkfrenkelone}\\
      0.7778 &  1728 &  113 &   (4) &    NIRM &\cite{pronkfrenkelone}\\
      0.7778 &   216 &  133 &   (3) &    ESPS &\cite{lshstwo}\\
      0.7778 &  1728 &  113 &   (3) &    ESPS &\cite{lshstwo}\\
      0.7778 &  5832 &  110 &   (3) &    ESPS &\cite{lshstwo}\\
      1.00   & 12000 &  260 & (100) &    NIRM &\cite{woodcock}\\
      1.0    &   216 &  131 &   (3) &    ESPS &\cite{lshstwo}\\
      1.0    &  1728 &  125 &   (3) &    ESPS &\cite{lshstwo}
    \end{tabular}
    \end{center}
\caption{
The difference in the {\em entropy} per particle
of the {\it fcc} and {\it hcp} crystalline phases of hard spheres;
the  associated uncertainties are in parenthesis. For comparison we note 
that the excess entropy per particle at 
$\rho / \rho_{cp} =0.7778$,  $N=1152$ is $-6.53\ldots$, with the phase-dependence showing in the 4th significant figure \protect\cite{frenkelladd}. 
}

\label{table:hardspheredata}
\end{table}


\begin{thebibliography}{20}
\bibitem{factorsofT} In the arguments of exponentials
we shall generally absorb factors of $kT \equiv 1/\beta$ 
into effective energy functions.
\bibitem{wellnotnaturally} It would be more circumspect to say that the 
partition function cannot in general be {\em naturally or usefully} 
expressed as a canonical average since that average inevitably falls into the 
category for which canonical sampling is inadequate \cite{badsampling}.
\bibitem{useofbigcon} Throughout this section we use the notation introduced in
Eq.~\ref{eq:bigcon} in which the phase label is absorbed into an
extended macrostate label $\bigcon$. The path defined by the set of
macrostates $\bigconset$ may (and sometimes will) extend from one
phase to another.
\bibitem{meaningofconjugate} We shall refer to a field $\lambda$ and a 
macrovariable $\plainop$ as conjugate if the $\lambda$-dependence of the 
configurational energy is of the form 
$\gene{\qset}{\smallcon, \lambda} = \gene{\qset}{\smallcon} -
\lambda  \plainop (\qset)$
\bibitem{seealsodepablo} 
Some of the results from \cite{nbwlj} are included
in the LJ phase diagram shown in Fig.~\ref{fig:depablotwo}.
\bibitem{trickypoint} 
Operationally we may regard a phase as `metastable' and assign it a
`free energy' (in relation to some other, also metastable) if the
conditions for single phase sampling (see Eq.~\ref{eq:qsetalphadefalt} and
accompanying discussion) are met.
\bibitem{funnygatewaystates} 
The configurations unearthed by the ES algorithm used in Lattice
Switch studies of hard spheres
(Section~\ref{subsubsec:switchexamples}) have some distinctive
features \cite{lshstwo}.
\bibitem{notjustesit}
The method to be discussed in the next section
belongs in this family; so does HPT, but less compellingly because of
the limitations we have already touched on in 
Section~\ref{subsubsec:paracpcritique}.
\bibitem{whynojacobeansbefore} In general the
partition function Eq.~\ref{eq:genpfdef} should already incorporate a 
Jacobean reflecting the transformation from cartesian coordinates to the
generalized coordinates $\qset$. It may always be absorbed into the 
configurational energy and we have adopted that convention.
\bibitem{constraintimplicit} The constraint is rarely made explict in MC 
studies: it is tacitly assumed that that the conditions for 
`single-phase sampling' (Eq.~\ref{eq:qsetalphadefalt} and accompanying
discussion) are met.
\bibitem{useoflattice} 
We use the term `lattice' rather loosely. The set of 
vectors $\Rset{\alpha}$ identifies the orthodox lattice convoluted with the
orthodox basis.
\bibitem{truealsoforcrystals} This will frequently be true for solid-solid 
phase behavior too; in the systems studied with Lattice Switch methods to 
date the effects have proved rather small.
\bibitem{carewithsets} 
We use  $\lambdaset$ to denote a set of physically 
different fields $\lambda_1, \lambda_2 \ldots$. In contrast $\lambdavalset$
(appearing in our discussion of paths) is shorthand for a set of 
different values $\lambda^{(1)},\lambda^{(2)} \ldots$ of 
some (particular) field $\lambda$. We use $\plainopset$ and $\plainopvalset$
in a similar way.
\bibitem{reducestocc} 
It reduces to the familiar form with the 
macrovariables $M_1$ and $M_2$ chosen as respectively the enthalpy and the 
volume whose conjugate fields are 1/T and $P/T$
\bibitem{equalareas} 
As an immediate corollary, the field values that 
satisfy the equal-areas criterion will identify the coexistence curve
of two such ensembles to within exponentially small corrections 
\cite{borgexperiment}.
\bibitem{useofE}
In this section we refer to  $E$ rather than its dimensionless 
counterpart $\genesimp \equiv \beta E$, since it seems  advisable to make 
the measure of temperature $\beta =1/kT$ explicit.
\bibitem{signconvention} Different authors have made different choices of 
notation and adopted different sign conventions at this point. We adopt those
of \cite{SMITH95}.
\bibitem{choiceofk} The constant $k$ merely serves to assign a convenient
lower (or upper) bound to the weights; the value chosen is absorbed
in the normalization of the sampling distribution.
\bibitem{caveat}
There is a caveat here: the traverse through the interesting region 
needs to be slow enough to allow a rough local equilibrium in each 
macrostate to be established; this condition is satisfied relatively poorly 
at the outset (where the MC trajectory heads rapidly for the 
equilibrium states); but it gets better the closer the sampling distribution 
comes to the multicanonical limit; one requires no more than this.
\bibitem{theconstant}
The unspecified constant in Eq.~\ref{eq:hrthree}
is prescribed by normalization.
\bibitem{hiddenparameters} We subsume into the
definitions of the macrovariables any parameters which do not have the 
credentials (`linearity') of the fields $\lambdaset$.
\bibitem{whyhistogram}
The name seems to reflect no more than the fact that the resulting
estimates of PDFs are as always based on 
(normalized) histograms accumulated  in the sampling process.
\bibitem{hrfromidentity}
The observable $\obs$ is then a  delta-function that picks out 
prescribed values for the 
macrovariables in the set $\plainopset$.
\bibitem{similarto}
To do so one must determine a set of weights that moderate the relative 
contributions of each histogram. 
There is more than a passing similarity 
here to the expanded ensemble framework discussed 
in Section~\ref{subsubsec:extendedsampling}.
\end{thebibliography}
\end{document}